\documentclass[acmsmall,nonacm, screen]{acmart}

\AtBeginDocument{%
  }

\usepackage{array}
\usepackage{xcolor}
\usepackage{multirow, lineno, pifont}
\usepackage{cleveref}
\usepackage[ruled, vlined, linesnumbered]{algorithm2e}

\definecolor{colortwo}{rgb}{0.4,0.77,0.17}
\definecolor{colorthree}{rgb}{0.01,0.51,0.93}

\newcommand{\dquote}[1]{``#1''}

\newcommand{\nc}{\newcommand}
\nc{\rnc}{\renewcommand}
\nc{\lbar}[1]{\overline{#1}}
\nc{\bra}[1]{\langle#1|}
\nc{\ket}[1]{|#1\rangle}
\nc{\ketbra}[2]{|#1\rangle\!\langle#2|}
\nc{\braket}[2]{\langle#1|#2\rangle}

\begin{document}

%%
%% The "title" command has an optional parameter,
%% allowing the author to define a "short title" to be used in page headers.
\title{Quantum Compiler Design for Qubit Mapping and Routing}
\subtitle{A Cross-Architectural Survey of Superconducting, Trapped-Ion, and Neutral Atom Systems}

%%
%% The "author" command and its associated commands are used to define
%% the authors and their affiliations.
%% Of note is the shared affiliation of the first two authors, and the
%% "authornote" and "authornotemark" commands
%% used to denote shared contribution to the research.
\author{Chenghong Zhu}
\authornote{Co-first Authors.}
\affiliation{%
  \institution{The Hong Kong University of Science and Technology (Guangzhou)}
  \city{Guangdong}
  % \state{Ohio}
  \country{China}
}

\author{Xian Wu}
\authornotemark[1]
\affiliation{%
  \institution{The Hong Kong University of Science and Technology (Guangzhou)}
  \city{Guangdong}
  % \state{Ohio}
  \country{China}
}

\author{Zhaohui Yang}
\affiliation{%
  \institution{The Hong Kong University of Science and Technology}
  % \city{HongKong}
  \country{Hong Kong}}

\author{Jingbo Wang}
\affiliation{%
  \institution{Beijing Academy of Quantum Information Sciences}
  \city{Beijing}
  \country{China}}

\author{Anbang Wu}
\affiliation{%
  \institution{Shanghai Jiao Tong University}
  \city{Shanghai}
  \country{China}}

\author{Shenggen Zheng}
\affiliation{%
  \institution{Quantum Science Center of Guangdong-Hong Kong-Macao Greater Bay Area}
  \city{Shenzhen}
  \country{China}}

\author{Xin Wang}
\authornote{Corresponding Author.}
\affiliation{%
  \institution{The Hong Kong University of Science and Technology (Guangzhou)}
  \city{Guangzhou}
  \country{China}}
\email{felixxinwang@hkust-gz.edu.cn}

% \author{John Smith}
% \affiliation{%
%   \institution{The Th{\o}rv{\"a}ld Group}
%   \city{Hekla}
%   \country{Iceland}}

%%
% \renewcommand{\shortauthors}{Trovato et al.}

%%
%% The abstract is a short summary of the work to be presented in the
%% article.
\begin{abstract}
Quantum hardware development is progressing rapidly with substantial advancements achieved across leading platforms, including superconducting circuits, trapped-ion systems, and neutral atom arrays. As the pursuit of practical quantum advantage continues, efficient quantum program compilation becomes essential for transforming high-level representations of quantum algorithms into physically executable circuits. A fundamental challenge in this process is qubit mapping and gate scheduling, which play a critical role in adapting compiled circuits to the architectural constraints and physical limitations of specific quantum hardware. In this survey, we systematically review and categorize research on the qubit mapping and routing problems across the three mainstream quantum hardware platforms. We primarily explore the development of hardware-aware compilers for superconducting platforms, classifying existing methods into solver-based, heuristic-based, and machine learning-based approaches, and analyze their optimization targets, including gate count, circuit duration, fidelity, and scalability. Furthermore, we examine the evolution of trapped-ion and neutral atom devices, analyzing the distinct challenges posed by their hardware characteristics and highlighting specialized compilers tailored to these unique physical constraints. Finally, we summarize the key challenges and identify some promising opportunities for future research in quantum compiler design across these hardware platforms.

\end{abstract}



%%
%% Keywords. The author(s) should pick words that accurately describe
%% the work being presented. Separate the keywords with commas.

% comment back
% \received{20 February 2007}
% \received[revised]{12 March 2009}
% \received[accepted]{5 June 2009}

%%
%% This command processes the author and affiliation and title
%% information and builds the first part of the formatted document.
\maketitle

\section{Introduction}

The rapid development of quantum computing has paved the way for groundbreaking advancements across diverse domains. From early theoretical breakthroughs demonstrating that algorithms such as Grover’s algorithm and Shor’s algorithm can outperform their classical counterparts, to the emergence of the Noisy Intermediate-Scale Quantum (NISQ) era~\cite{Preskill2018}, the field of quantum computing has progressed rapidly toward the realization of Fault-Tolerant Application-Scale Quantum (FASQ) computers~\cite{preskill2025beyond}.
The viability of quantum computing has been convincingly demonstrated across several hardware platforms, including superconducting circuits~\cite{nakamura1999coherent}, Rydberg atoms~\cite{lukin2001dipole}, trapped-ions~\cite{cirac1995quantum}, photonics~\cite{knill2001scheme}, and silicon quantum dots~\cite{pla2012single}. Notably, quantum advantage has already been experimentally shown in tasks like random circuit sampling~\cite{Arute2019,Zhong2020_z,wu2021strong} and learning the properties of physical systems~\cite{Huang_2022}. These accomplishments, both in algorithm design and experimental hardware, offer strong evidence that quantum computing can drive transformative solutions to diverse applications, ranging from drug discovery~\cite{cao2018potential} and tackling complex financial problems~\cite{orus2019quantum} to advances in artificial intelligence~\cite{dunjko2018machine}.

To achieve the practical quantum advantage, coordinated efforts are required across multiple layers of the quantum computing technical stack. Although theoretical quantum algorithms are frequently devised at the logical level, they cannot be executed directly on hardware due to constraints such as limited coherence times, varying native gate sets, and diverse physical qubit connectivities, across different platforms~\cite{Linke_2017}. Consequently, it is necessary to perform and effective conversion process---\dquote{quantum program compilation} that translates high-level quantum algorithms into low-level quantum instructions that adhere to each device’s specific requirements.

\textit{Quantum program compilation} or \textit{quantum compilation}~\cite{ding2022quantum, zulehner2020introducing} involves multiple stages, including transforming quantum algorithms into unitary operations, decomposing gates into native sets supported by specific devices, and adapting the resulting operations to hardware constraints. A key step in this process is \textit{qubit mapping and gate scheduling}, which ensures that the compiled circuits remain compatible with the targeted device. However, these hardware constraints vary among mainstream platforms. For instance, superconducting devices often require additional SWAP operations to handle limited connectivity~\cite{li2019tackling}, whereas trapped‐ion devices rely on ion shuttling~\cite{murali2020architecting_iontrap}. Such extra operations can impose significant overhead, underscoring the need for device‐dependent strategies that minimize these costs and thus guarantee more reliable execution of the compiled circuits.

\begin{figure}[t]
    \centering
    \includegraphics[width=1\linewidth]{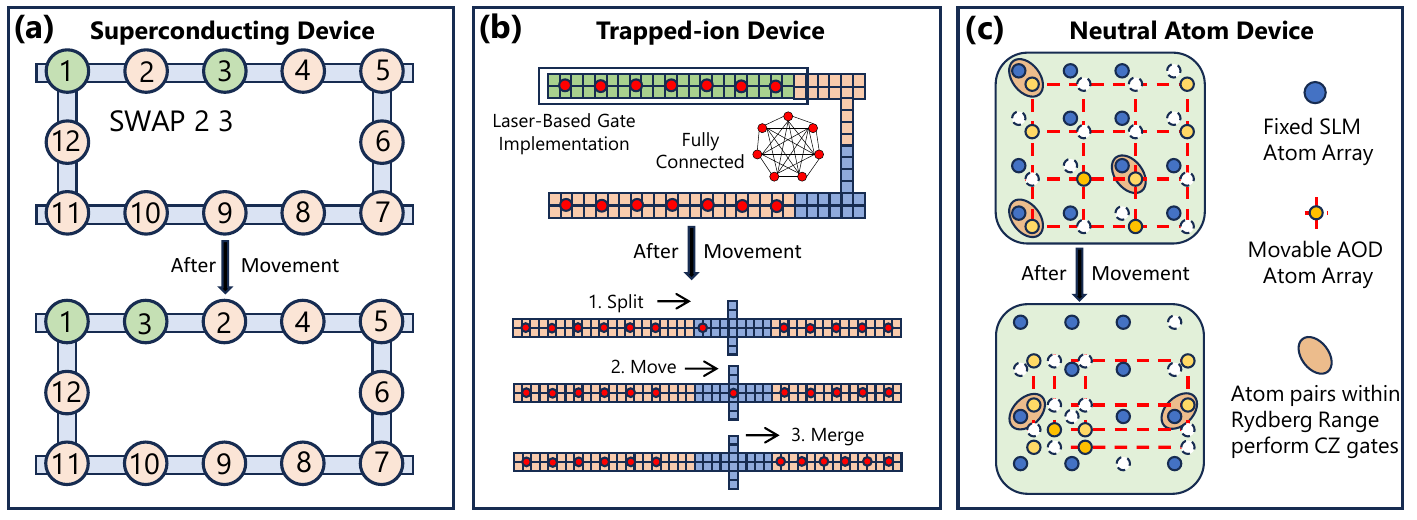}
    \caption{ Overview of the three mainstream quantum hardware architectures and their graph illustrations: (a) superconducting devices, (b) trapped-ion devices, and (c) neutral atom devices.
    }
    \label{fig:different_platform}
\end{figure}

Fig.~\ref{fig:different_platform} presents an overview of the three mainstream quantum hardware architectures that have seen rapid development in recent years: superconducting devices, trapped-ion devices, and neutral atom devices, along with their respective coupling map abstractions.
This progress is mainly due to the demonstrated versatility of the gate sets they can implement and significant advances in scalability and accuracy. The superconducting platform, with its mature fabrication process and long history, can produce chips with diverse topological structures~\cite{krantz2019quantum, kjaergaard2020superconducting}. However, challenges such as locally connected topologies and severe qubit crosstalk must be addressed~\cite{mundada2019suppression, tripathi2022suppression, zhao2022quantum}. The trapped-ion platform excels in extremely high precision in quantum gate operations~\cite{harty2014high, ballance2016high, smith2024single}. It ensures full connectivity within a single trap~\cite{bruzewicz2019trapped}, yet its scalability and gate operation parallelism still require shuttle-based Quantum Charge Coupled Device (QCCD) architectures~\cite{pino2021demonstration, kaushal2020shuttling}. In contrast, the neutral atom platform effectively overcomes trapped-ion devices' scalability and parallelism limitations~\cite{saffman2019quantum,evered2023high}. Regarding quantum compilation, key aspects such as qubit mapping and gate scheduling are primarily influenced by gate fidelity, qubit connectivity, and parallelism. Gate fidelity determines algorithm reliability, connectivity impacts qubit mapping efficiency, and parallelism constrains gate scheduling speed. Research into compilation strategies for these platforms focuses on optimizing circuit execution under hardware-specific constraints and offers transferable methodological approaches that can be adapted to other quantum systems, supporting the development of a universal quantum software stack.

To evaluate the performance of qubit mapping and gate scheduling algorithms, we can broadly categorize the relevant metrics into two classes: \textit{quantum algorithm performance} and \textit{classical resource consumption}. This distinction reflects the dual objective of optimizing both the execution quality of quantum circuits and the efficiency of the classical compilation process. In terms of quantum algorithm performance, we track metrics such as \textit{gate count}, \textit{execution time}, \textit{circuit depth}, and ultimately \textit{fidelity}. Although superconducting, trapped‐ion, and neutral‐atom platforms differ in coherence times, noise characteristics, and connectivity constraints, these metrics remain central for comparing algorithmic outcomes across diverse hardware implementations. In terms of classical resource consumption, as the field transitions from NISQ systems toward fault‐tolerant quantum computing (FTQC), the rapid growth in qubit numbers introduces substantial scalability challenges. Designing compilation algorithms that can handle large‐scale circuits efficiently within reasonable computational time and memory costs becomes crucial for ensuring timely results and effectively harnessing the increasing capabilities of emerging quantum devices.

\textbf{Contributions.} In summary, the contributions of this survey paper are as follows: 
\begin{itemize}
    \item We systematically analyze the hardware constraints of three  kinds of mainstream quantum computing architectures---superconducting, trapped-ion, and neutral atom systems.
    \item We organize and categorize the major qubit mapping and gate scheduling algorithms in the context of these three kinds of hardware platforms.
    \item We propose several forward-looking research directions that may serve as useful guidance for future developments in quantum compiler design.
\end{itemize}

\begin{table}[t]
    \centering
    \caption{Content Guidance of This Paper}
    \small % Reduce font size
    \begin{tabular}{|c|cc|}
        \hline
        \textbf{Section} & \textbf{Subsection} & \textbf{Index} \\
        \hline
        \multirow{3}{*}{Background} & Quantum Computation & \S\ref{subsec:quantumcomputation} \\
        & Quantum Program Compilation & \S\ref{subsec:circuit_compilation_pipeline} \\
        & Quantum Hardware & \S\ref{subsec:qhardware} \\
        \hline
        \multirow{4}{*}{Superconducting} & Solver-Based Compiler & \S\ref{subsec:solver} \\
        & Heuristic-Based Compiler & \S \ref{subsec:heuristic} \\
        & AI-Based Compiler & \S\ref{subsec: ai sc} \\
        & Application-Specific Compilers & \S \ref{subsec:application_specific} \\
        \hline
        \multirow{3}{*}{Trapped-Ion} & Linear Tape Device & \S\ref{subsec:linear tape} \\
        & Quantum Charge-Coupled Device & \S \ref{subsec:qccd}\\
        & Large Scale Distributed Device & \S \ref{subsec:large scale trapped-ion}\\
        \hline
        \multirow{3}{*}{Neutral Atom} &Fixed Neutral Atom Arrays  & \S\ref{subsec: early na} \\
        & Reconfigurable Neutral Atom Devices & \S\ref{subsec: adavanced na} \\
        & Zoned Architecture Neutral Atom Devices & \S\ref{subsec: zoned na} \\
        \hline
        \multicolumn{2}{|c}{Future Direction and Conclusion} & \S~\ref{sec:conclusion} \\
        \hline
    \end{tabular}
    \label{table:organization}
\end{table}

\textbf{Organization.} The rest of this paper is structured as follows. Section~\ref{sec:pre} begins with an introduction to quantum computation, then transitions to an overview of three mainstream quantum hardware and delves into the introduction of quantum circuit compilations. In Section~\ref{sec:superconducting compiler}, we focus on the compilers designed for superconducting chips. Then we provide a comprehensive overview of trapped-ion devices, tracing the compiler evolution from single-trap to multi-trap architectures, and ultimately to distributed large-scale trapped-ion computing systems in Section~\ref{sec:trapped-ion}. Section~\ref{sec:neutral atom} focuses primarily on recently developed neutral atom devices and will discuss compilers designed to address the unique features and constraints of these devices. We conclude and provide the future direction in Section~\ref{sec:conclusion}. The organization of this survey is shown in Table~\ref{table:organization}.

\section{Background}
\label{sec:pre}

\subsection{Quantum Computation}\label{subsec:quantumcomputation}
\textbf{Qubit.} In quantum computing, the basic unit of quantum information is the quantum bit or qubit, whose state is described and evolves in accordance with the framework of quantum mechanics. A single-qubit pure state is described by a unit vector in the Hilbert space $\mathcal{H}(\mathbb{C}^2)$, which is commonly written in Dirac notation $\ket{\psi} = \alpha\ket{0} + \beta\ket{1}$, with $\ket{0} = (1,0)^T$, $\ket{1} =(0,1)^T$ and $\alpha,\beta\in \mathbb{C}$ subject to $|\alpha|^2 + |\beta|^2 = 1$. The complex conjugate of $\ket{\psi}$ is denoted as $\bra{\psi} = \ket{\psi}^\dagger$. The $N$-qubit Hilbert space is formed by the tensor product ``$\otimes$'' of $N$ single-qubit state spaces with dimension $d=2^N$. For an $N$-qubit state $\ket{\psi}\in \mathcal{H}(\mathbb{C}^{d})$, it can be written as $\ket{\psi} = \sum_{i=0}^{d-1} a_i \ket{i}$, with $\sum_{i=0}^{d-1} \lvert a_i\rvert^2 = 1$ and $\{\ket{i}\}_{i=0}^{d-1}$ form a set of orthogonal basis vectors, each $\ket{i}$ having 1 in the $i$-th element and 0 elsewhere.

\textbf{Quantum Operation.} In the representation of quantum circuit model, quantum computing performs by altering the states of qubits through quantum operations (or gates). Similar to classical computing, QC systems generally support a universal gate set usually composed of basic single-qubit rotations and one or more two-qubit gates. These gates are capable of acting on one or two qubits simultaneously and expressing any quantum program~\cite{Barenco_1995}. Common single-qubit rotation gates include $R_x(\theta)=e^{-i\frac{\theta}{2}X}$, $R_y(\theta)=e^{-i\frac{\theta}{2}Y}$, $R_z(\theta)=e^{-i\frac{\theta}{2}Z}$, which are in the matrix exponential form of Pauli matrices,
\begin{equation}
    X = \begin{pmatrix}
        0 & 1 \\ 1 & 0
    \end{pmatrix},\quad
    Y = \begin{pmatrix}
        0 & -i \\ i & 0 \\
    \end{pmatrix},\quad
    Z = \begin{pmatrix}
        1 & 0 \\ 0 & -1 \\
    \end{pmatrix}.
\end{equation}
Common two-qubit gates in superconducting devices include controlled-X gate $\text{CNOT} = \ketbra{0}{0} \otimes I + \ketbra{1}{1} \otimes X$  and controlled-Z gate $\text{CZ}= \ketbra{0}{0} \otimes I + \ketbra{1}{1} \otimes Z$, which can generate quantum entanglement among qubits. Parametrized $XX(\theta)=e^{-i\frac{\theta}{2} X\otimes X}$ gates and CZ gate are the most frequently used two-qubit gates in trapped-ion devics and neutral atom devices, respectively. SWAP gate swaps the states of two qubits and can be decomposed into three CNOT gates.

\textbf{Graph representation of quantum circuit.} Considering that a quantum circuit is fundamentally a time-ordered series of quantum instructions, it is convenient to adopt the formal graph representation to illustrate the causal relationships between these instructions. This is known as the Directed Acyclic Graph (DAG) representation of quantum circuits. 
Within the DAG of a quantum circuit, each quantum operation is represented by a vertex in the graph, and a directed edge from vertex $g_i$ to vertex $g_j$ signifies that the quantum operation at vertex $g_i$ must precede the operation at vertex $g_j$.
The presence of any cycle within the graph would imply that past instructions are dependent upon future ones, thereby violating the causal structure of quantum circuits. Consequently, this directed graph is inherently acyclic.

\begin{figure}[H]
    \centering
    \includegraphics[width=0.6\linewidth]{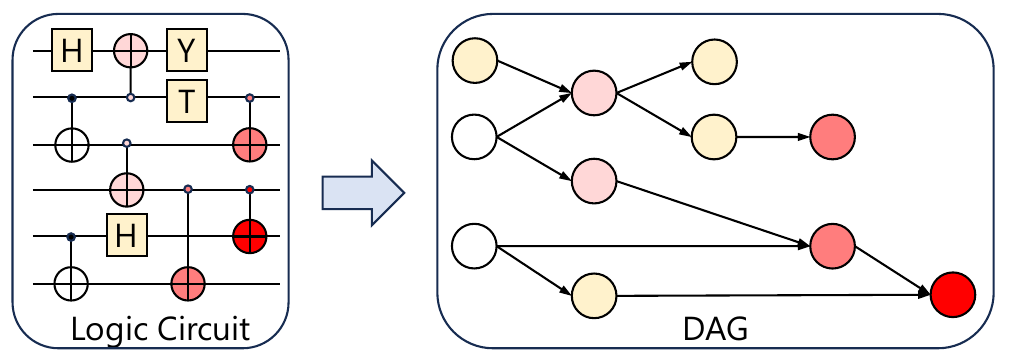}
    \caption{An illustration of a circuit's DAG representation. 
    The front layer of the graph is made up of the nodes with an in-degree of 0, these gates can be executed immediately. A gate is removed from the dependency graph after it has been executed, which causes some gates to enter the front layer.
    }
    \label{fig:dependency graph}
\end{figure}

Fig.~\ref{fig:dependency graph} provides an example of DAG representation. DAG representations are primarily employed to represent dependencies between quantum gates.
In a DAG, nodes with an in-degree of zero indicate that the corresponding gates can be executed immediately, provided that the relevant physical device constraints are satisfied. Consequently, to streamline the representation, researchers typically retain only the edges that directly connect each gate to its immediate predecessors.
The nodes representing single-qubit gates within the DAG are observed to have no impact on the dependency relationships among the nodes. Therefore, these nodes are generally omitted during the construction of the DAG to streamline the representation and analysis of the underlying quantum circuit structure.

\subsection{Quantum Program Compilation}\label{subsec:circuit_compilation_pipeline}

\textbf{Overview.} One of the most challenging obstacles to running computational applications such as machine learning and chemical simulations is to lower quantum algorithms to physically executable circuits (gate sequences) that align with specific architecture constraints of quantum hardware~\cite{kusyk2021survey,huang2023near}. This transformation process is referred to as \emph{quantum program compilation} or \emph{Quantum Compilation}~\cite{ding2022quantum, zulehner2020introducing}. Although there are differences in specific definitions, it can be considered that quantum program compilation includes the following tasks:

\begin{enumerate}
    \item \textbf{Circuit Synthesis}.  For a given algorithm $\mathcal{A}$ and a given native gate library $\mathcal{L}$, one should output a circuit (gate sequence) $\mathcal{C}$ to implement $\mathcal{A}$ and each gate in $\mathcal{C}$ is from $\mathcal{L}$. Notably, $\mathcal{A}$ may be given in different forms, for example, unitary matrices, gate sequences, Hamiltonians, and so on.
    \item \textbf{Optimization (logical level)}.
    For a given circuit $\mathcal{C}$, one can optimize the circuit to obtain a less costly circuit $\mathcal{C'}$ (e.g., fewer gate count, less circuit depth). This process is generally at the logical level without considering the corresponding physical hardware.

    \item \textbf{Qubit Mapping and Gate Scheduling (M\&S)}. 
    For a given circuit $\mathcal{C'}$ and a given hardware constraint, one should apply proper operation to generate a new circuit $\mathcal{C''}$ that satisfies the hardware constraint while incurring the lowest possible cost.
    
\end{enumerate}

\begin{figure}[h]
     \centering
     \includegraphics[width=1.0\linewidth]{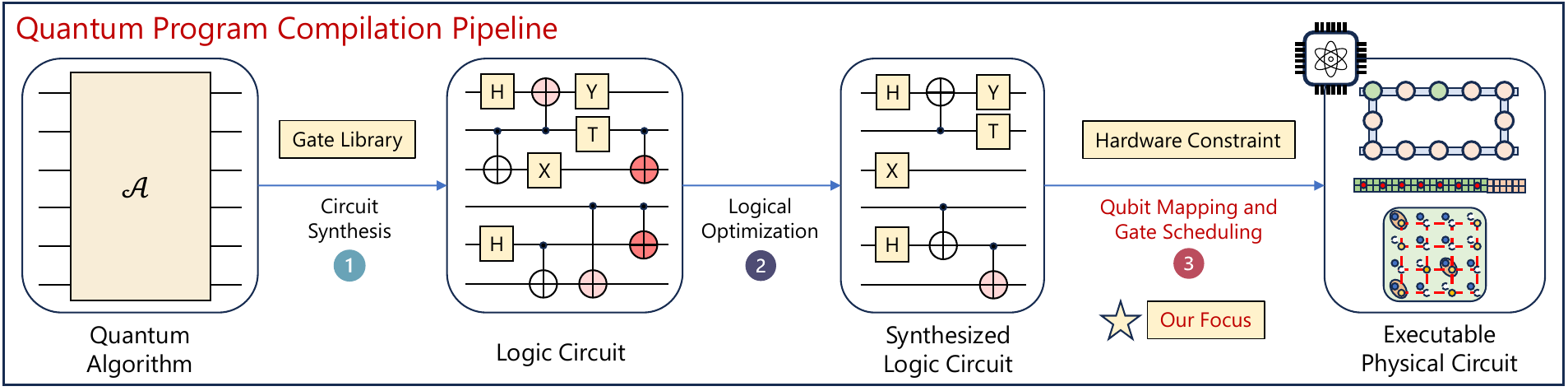}
     \caption{The workflow of quantum program compilation.}
     \label{fig:compilation framework}
\end{figure}
The typical workflow of quantum program compilation is shown in Fig.~\ref{fig:compilation framework}, although there might be alternative compilation workflows (e.g., Qiskit focuses on post-mapping optimization for physical circuits). Circuit synthesis and optimization are typically taken into consideration simultaneously. As these processes are typically platform-independent, hardware constraints are not necessarily considered. 

\textbf{Circuit Synthesis and Logical Level Optimization. } 
For the decomposition of general unitary matrices, \citet{shende2005synthesis} proposed the Quantum Shannon Decomposition, which achieves asymptotic optimality in the number of CNOT gates required. Subsequent works have continued to refine and extend this approach~\cite{krol2022efficient, krol2024beyond}. However, in practical applications, the focus has increasingly shifted toward the decomposition of specific classes of unitary matrices, such as those representing Boolean functions~\cite{zi2025shallow, yang2023efficient, wu2024asymptotically}, where tailored methods can yield more efficient circuit realizations.

Currently, there are two common methods for implementing quantum circuits for Boolean functions. One approach is based on the Exclusive Sum of Products (ESOP) representation \cite{fazel2007esop}. Since a product term naturally corresponds to a multi-controlled Toffoli (MCT) gate in quantum circuits, this approach was widely adopted in earlier research.
However, ESOP is not efficient enough to optimize the utilization of quantum computing resources in the NISQ era, and the overhead of MCT gates is generally considered to grow exponentially with the number of control qubits \cite{mottonen2004quantum}. Moreover, extensive research has been conducted in this area by Giulia Meuli et al.  They first proposed a heuristic algorithm based on XAG and demonstrated that the number of auxiliary qubits can be bounded by the number of AND gates in the XAG \cite{meuli2019role}.  In addition, they proposed algorithms for balancing the trade-off between the number of quantum bits and quantum operations, modeling it as instances of reversible pebble games \cite{meuli2019reversible}.  Finally, they further proposed three XAG-based algorithms for synthesizing quantum circuits in Clifford + T libraries, aiming to minimize the number of T gates, the T-depth, and the number of qubits, respectively \cite{meuli2022xor}. In addition to these approaches, recent studies have continued to advance the decomposition of unitary matrices. For a comprehensive review of circuit synthesis and decomposition techniques, readers are referred to \citet{ge2024quantumcircuitsynthesiscompilation}.

\textbf{Qubit Mapping and Gate Scheduling. } In contrast, the \textbf{M\&S} task is closely tied to hardware limitations. As a result, most hardware-specific compilers prioritize the \textbf{M\&S} task.
In general, the \textbf{M\&S} problem has two parts, one is gate scheduling and the other is initial mapping. The gate scheduling part refers to the part mentioned above, in short, is how to insert proper operations. The initial mapping is to assign  qubits at the beginning such that the overhead of gate scheduling is minimal. It has been proved that the initial mapping can significantly affect the final circuit quality \cite{siraichi2018qubit, zulehner2018efficient}, which means one need to consider how to give a high-quality initial mapping. However, it also suggests that the overall cost will be dominated by the scheduling parts as the circuit depth increases along with its gate count \cite{cheng2024robust}, which means the initial mapping will not be so important in large-scale circuits. Several key metrics define the optimization performance of M\&S task and the objectives of these compilers can be categorized as follows:
\begin{enumerate}
    \item \textbf{Gate Count}. Since the gates are not ideal on NISQ devices, an increased number of gates correlates with decreased fidelity. Additionally, quantum gate implementation incurs certain costs. This facet represents a crucial avenue in quantum computing research, focusing on developing more practical and efficient quantum algorithms and applications.
    \item \textbf{Circuit Depth}. Often aligned with the concept of circuit parallelization, this research aspect strives to minimize circuit depth for a specified functionality. Circuit depth is intrinsically linked to the execution time of a quantum circuit.
    \item \textbf{Fidelity}. 
    Due to inherent imperfections in real quantum circuits, fidelity is also commonly used to quantify the similarity between a circuit's actual output and its ideal counterpart. The evaluation of circuit fidelity is influenced by specific hardware parameters and is typically analyzed within the context of practical applications.
    \item \textbf{Compilation time}. As hardware advances, the need for scalable compilers becomes essential. Compilation time refers to the time required to transform a logical circuit into a physically executable one.
\end{enumerate}

\subsection{Quantum Hardware}\label{subsec:qhardware}
Recent years have witnessed remarkable advances in quantum hardware based on superconducting circuits, trapped-ions, and neutral atom arrays. Table~\ref{tab:hardware-properties} summarizes representative performance metrics for these three platforms. Superconducting processors achieve the shortest single‑ and two‑qubit gate times, typically on the order of tens of nanoseconds, whereas trapped‑ion systems operate several orders of magnitude more slowly. The extended interaction times in trapped‑ion and neutral‑atom devices, however, support longer coherence times and correspondingly higher gate fidelity.

\begin{table}[h]
    \centering
    \caption{Characteristics of three quantum hardware}
    \begin{tabular}{|c|c|c|c|}
    \hline
         Hardware&  Trapped-ion & Superconducting & Neutral Atom\\
         \hline
         gate time& $1 \sim 100 \mu s$ & $10 \sim 100 ns$ & $0.2 \sim 5 \mu s$ \\
         \hline
         connectivity &full&partial &partial \\
         \hline
         physical movement & supported & unsupported & supported\\
         \hline
         decoherence time &$5500 s$\cite{mei2022experimental} & $300 \mu s$\cite{place2021new} & 40$\pm7s $\cite{barnes2022assembly}\\
         \hline
         1Q gate fidelity &  99.9934(3)\% \cite{ballance2016high} & $\approx $99.9\% \cite{xu2020high} &99.98\%\cite{graham2022multi_atom}  \\
         \hline
         2Q gate fidelity &99.9(1)\% \cite{ballance2016high} & 99.5\% \cite{xu2020high} &99.5\%\cite{evered2023high} \\
         \hline
         Parallelism & Low & Moderate & High \\
         \hline
    \end{tabular}
    \label{tab:hardware-properties}
\end{table}

Qubits can be constructed via two fundamentally different approaches. The first follows a \textit{top-down paradigm}, wherein engineered quantum states are fabricated using mature CMOS technologies. Superconducting qubits, a representative of this approach, exemplify solid-state platforms where qubit control and integration are enabled by advanced nanofabrication. The second is a \textit{bottom-up approach}, which leverages naturally occurring quantum systems as qubit carriers. Since the physical world is inherently quantum at its core, atoms and ions offer stable and coherent quantum states that are ideal for quantum information processing. Trapped-ions and neutral atoms are the most prominent platforms in this category.

A central challenge in quantum compilation is to map abstract quantum algorithms onto noisy, physically constrained hardware for efficient execution. This task requires addressing limitations in \textbf{connectivity} and \textbf{parallelism}, both of which vary significantly across quantum hardware platforms. Notably, superconducting circuits, neutral atom arrays, and trapped-ion systems together span the full spectrum of compilation challenges, making them ideal case studies.

In terms of \textit{qubit connectivity}, three categories can be identified. (1) \textit{Local connectivity}: Superconducting qubits exhibit fixed and local connectivity, determined at the fabrication stage. Qubit interactions are typically limited to nearest neighbors, making qubit mapping and SWAP gate insertion major bottlenecks in the compilation process. (2) \textit{Reconfigurable connectivity}: Neutral atom systems enable dynamic connectivity via atom rearrangement and selective excitation. Compilation must jointly optimize atom placement and gate scheduling, accounting for movement costs and Rydberg blockade constraints. (3) \textit{All-to-all connectivity}: Trapped-ion systems support fully connected architectures through global laser addressing, allowing arbitrary entangling operations. This reduces mapping complexity but shifts focus toward optimizing gate parallelization, managing phonon modes, and minimizing laser-induced crosstalk.

Similarly, qubit \textit{parallelism} can be categorized into three regimes. (1) \textit{Low parallelism}: In trapped-ion platforms, Coulomb interactions are long-range and collective, making it difficult to execute multiple two-qubit gates in parallel within a single trap. Limited parallelism can be achieved using multi-zone architectures such as QCCD. (2) \textit{Moderate parallelism}: Superconducting systems allow some degree of parallelism, but gate operations between nearby qubits often suffer from crosstalk. Compilation strategies must carefully schedule gate operations to avoid interference. (3) \textit{High parallelism}: Neutral atoms offer the highest degree of parallelism. Optical tweezers can spatially isolate gate operations, enabling simultaneous execution of many two-qubit gates without crosstalk. Here, the main challenge lies in optimizing atom movement to minimize transport overhead.

We focus on these three hardware platforms not only because they represent the most rapidly advancing and promising candidates for scalable, fault-tolerant quantum computing, but also because their distinct physical characteristics comprehensively cover the key compilation challenges of connectivity, parallelism, and qubit modality. Understanding and addressing compilation strategies across these platforms will offer valuable insights transferable to emerging quantum architectures.

\section{Superconducting compiler}
\label{sec:superconducting compiler}

\subsection{Background}
Superconducting hardware, particularly in the context of quantum computing, involves designing and utilizing circuits that operate at extremely low temperatures where certain materials exhibit superconductivity - the property of zero electrical resistance. These circuits typically consist of Josephson junctions, which are the fundamental building blocks for superconducting qubits. 

The properties of superconducting systems are less complex compared to those of trapped-ion and neutral atom systems, allowing researchers to primarily focus on qubit connectivity. In superconducting devices, each physical qubit can only directly couple with its neighboring qubits. As a result, for a given mapping, two-qubit gates can only be applied to specific logical qubit pairs whose corresponding physical qubit pairs support direct coupling.

\textbf{Constraints.} In superconducting devices, two-qubit gates can only be applied to qubit pairs that are physically connected. The topology of qubits in a quantum device is typically fixed and can be formally represented by a coupling graph $G_q=(Q, E_q)$, where $Q$ denotes the set of qubits and $E_q$ specifies the allowable two-qubit interactions. While single-qubit gates can be applied to any qubit in $Q$, two-qubit gates are restricted to qubit pairs $(q_i, q_j)$ such that \((q_i, q_j) \in E_q\).

\begin{figure}[h]
    \centering
    \includegraphics[width=0.95\linewidth]{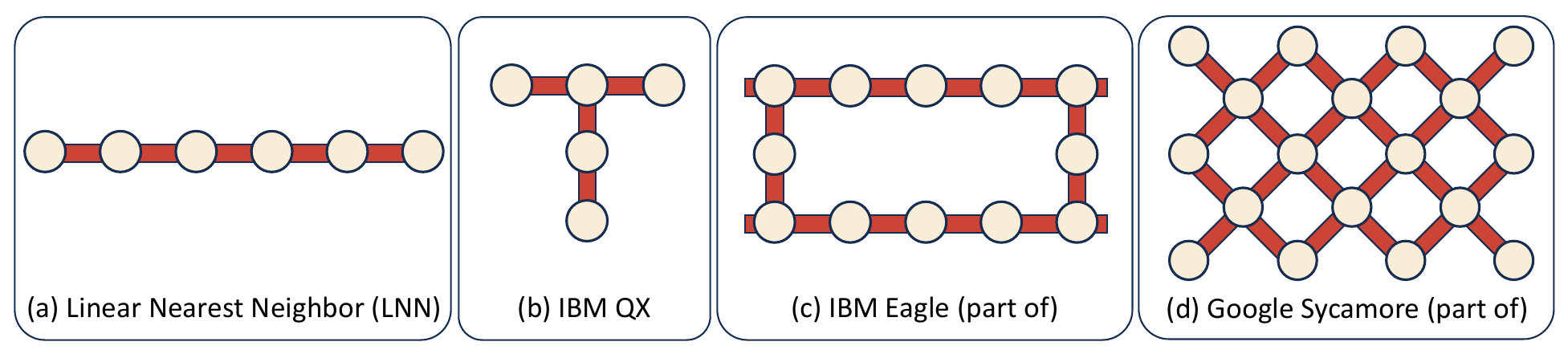}
    \caption{Common structures of superconductivity devices and their corresponding connectivity graphs.}
    \label{fig:common structure}
\end{figure}

Fig.~\ref{fig:common structure} shows the coupling graph of some common superconducting architecture. Each vertex represents a physical qubit, and each edge represents the connectivity between two qubits. A two-qubit gate is allowed to act on $(q_i,q_j)$ if and only if $(q_i,q_j) \in E_q$. If $(q_i,q_j) \not\in E_q$, researchers need to insert SWAP gates to make the logical qubits adjacent to each other in the physical circuit. To execute a generic quantum circuit (with arbitrary interactions between its qubits) on such an architecture, the circuit needs to be mapped. This involves qubit allocation, where the circuit’s (logical) qubits are assigned to the device’s (physical) qubits in an initial layout and routing. The original circuit is augmented with SWAP gates such that it adheres to the target device’s coupling graph. The Qubit Mapping and Gate Scheduling (M\&S) problem—also referred to as Qubit Mapping and Routing—is most commonly studied in the context of superconducting devices. It is known that determining whether a given gate sequence can be implemented using no more than $K_s$ SWAP gates is NP-complete~\cite{siraichi2018qubit}, and finding an optimal qubit mapping for a quantum circuit has been proven to be NP-hard~\cite{botea2018complexity}.

\textbf{Problem Definition: } For a given quantum circuit and the coupling graphs of superconducting chips, the task is to devise an initial mapping of logical qubits to physical qubits and insert proper SWAP gates to ensure that the result circuit satisfies the constraints imposed by two-qubit gate interactions.

\begin{table*}[t]
\centering
\caption{Quantum compilation methods and their target metrics, along with proposed approaches for superconducting devices.}
\resizebox{1.0\linewidth}{!}{
\begin{tabular}{c|c||c|c|c|c||c|c|c|c|c||c}
\toprule
\multirow{3}{*}{\textbf{Ref.}} &\multirow{3}{*}{\textbf{Year}} & \multicolumn{4}{c||}{\textbf{Target}}& \multicolumn{5}{c||}{\textbf{Methods}} & \multirow{3}{1.6cm}{\centering \textbf{(Near)} \textbf{Optimal?}} \\
\cline{3-11}
 & &\textbf{Gate} &\textbf{Execution} &\multirow{2}{*}{\textbf{Fidelity}}  & \multirow{2}{*}{\textbf{Others}}  & 
\multirow{2}{*}{\textbf{SMT}} & \multirow{2}{*}{\textbf{SAT}} & \multirow{2}{*}{\textbf{ILP}} & \multirow{2}{*}{\textbf{Heuristic}} & \multirow{2}{*}{\textbf{AI}} &  
 \\
 & &\textbf{Count}&\textbf{Time} & & & & & & &
 \\
\hline
\cite{bhattacharjee2017depth} &2017& & \checkmark& & & &  &\checkmark & & &\checkmark 
\\
\hline
\cite{meuli2018sat} & 2018 &\checkmark& &  & & &\checkmark & & & &\checkmark
\\
\hline
\cite{zulehner2018efficient} & 2018 &\checkmark& &  & & & & &\checkmark & &\checkmark
\\
\hline
\cite{oddi2018greedy} & 2018 & & &  & \checkmark& & & &\checkmark & &\ding{55}
\\
\hline
\cite{murali2019noise} &2019& & \checkmark &\checkmark && \checkmark & & & & & \checkmark
\\
\hline
\cite{tannu2019not} &2019& & &\checkmark& &  & & & \checkmark& & \ding{55}
\\
\hline
\cite{wille2019mapping}&2019 &\checkmark& &  & & &\checkmark & & & &\checkmark
\\
\hline
\cite{bhattacharjee2019muqut}&2019 &\checkmark & \checkmark& & & &  &\checkmark & & &\checkmark
\\
\hline
\cite{li2019tackling}&2019 & \checkmark & & & & & & & \checkmark & & \ding{55}
\\
\hline
\cite{tan2020optimal}& 2020  & \checkmark&
  &  &  &  \checkmark   &  &     % Circuit Repr
  &  &  & \checkmark % Other columns
\\
\hline
\cite{niu2020hardware}& 2020  &\checkmark 
  &  & \checkmark & &    &  &     % Circuit Repr
  & \checkmark  &  & \ding{55} % Other columns
\\
\hline
\cite{zhu2020dynamic}& 2020  &\checkmark &
  &  & &    &  &     % Circuit Repr
  & \checkmark  &  & \ding{55} % Other columns
\\
\hline
\cite{zhu2021iterated}&2021 & \checkmark &  & & & & & &\checkmark  & & \ding{55}
\\
% \hline
% \cite{fosel2021quantum}&2021 & \checkmark & \checkmark& & & & & & & \checkmark & \ding{55}
% \\
\hline
\cite{zhang2021time}&2021 & & \checkmark& & & & & & \checkmark & & \checkmark
\\
\hline
\cite{patel2022quest}&2022 &\checkmark & & \checkmark& & & & &\checkmark & &\ding{55}
\\
\hline
\cite{huang2022reinforcement}&2022 & \checkmark &&  & & & & & & \checkmark & \ding{55}
\\
\hline
\cite{sinha2022qubit} &2022& &  \checkmark& & & & & & & \checkmark & \ding{55}
\\
\hline
\cite{liu2022not} &2022&   \checkmark &\checkmark && & & & & \checkmark & & \ding{55}
\\
\hline
\cite{park2022fast} &2022&   \checkmark & & &\checkmark & & & & \checkmark & & \ding{55}
\\
\hline
\cite{schneider2023sat}&2023 &\checkmark& &  & & &\checkmark & & & &\checkmark
\\
\hline
\cite{paler2023machine} &2023& &  \checkmark& & & & & & & \checkmark & \ding{55}
\\
\hline
\cite{niu2023enabling} &2023&\checkmark &  \checkmark &\checkmark & & & & &  \checkmark & & \ding{55}
\\
\hline
\cite{zhu2023variation} &2023&\checkmark &    & \checkmark  & & & & & \checkmark & & \ding{55}
\\
\hline
\cite{fu2024effective} &2024& &  \checkmark && & & & &  \checkmark & &  \checkmark
\\
\hline
\cite{yu2024symmetry} &2024& &  & &\checkmark & & & &  \checkmark & & \ding{55}
\\
\hline
\cite{steinberg2024lightcone} &2024&\checkmark &  & &\checkmark & & & &  \checkmark & & \ding{55}
\\
\hline
\cite{escofet2024route} &2024& &\checkmark  & &\checkmark & & & &  \checkmark & & \ding{55}
\\
\hline
\cite{cheng2024robust} &2024&\checkmark &  & & & & & &  \checkmark & & \checkmark
\\
\hline
\cite{quetschlich2024towards} &2024&\checkmark & \checkmark &\checkmark &\checkmark & & & &   &\checkmark & \ding{55}
\\
\hline
\cite{pascoal2024deep} &2024& &  &\checkmark & & & & &   &\checkmark & \ding{55}
\\
\bottomrule
\end{tabular}
}
\label{tab:superconducting full table}
\end{table*}
In the context of superconducting devices, various methods have been proposed to minimize the additional costs incurred when transforming quantum circuits into hardware-compliant configurations. 
We categorize these methods into three groups: \textbf{solver based} for (near) optimal solutions, \textbf{heuristic algorithms}, and \textbf{machine-learning based} approaches. It is noted that the survey conducted by~\citet{kusyk2021survey} primarily addresses superconducting quantum compilation methods that leverage artificial intelligence (AI) and heuristic strategies. This survey will focus on scalable heuristic-based methods that are utilized in industrial-level compilers for superconducting devices, highlighting advancements in state-of-the-art compilers since 2021.
We summarize the key developments and representative works in this area in Table \ref{tab:superconducting full table}.

\subsection{Solver-Based Compiler}
\label{subsec:solver}
Due to the fact that the \textbf{M\&S} problem associated with superconducting circuits can be accurately depicted through mathematical models, numerous efforts are directed towards modeling these circuits. Subsequently, solvers are employed to find solutions. We refer to this approach as the solver-based method.
The advantage of this method lies in its ability to find the optimal solution (or an approximation of it). However, the drawback is that the models used are mostly NP-hard, meaning that the computational time can be excessively long.
Next, we will reveal a diverse array of strategies that leverage mathematical formulations and solver software. 

Notably, Satisfiability Modulo Theories (SMT) solvers have been widely adopted for their ability to find optimal solutions through exact methods, albeit at the cost of exponential computational time. For instance, \citet{lye2015determining} employ Pseudo-Boolean Optimization (PBO) in the context of multi-dimensional circuits, while \citet{tan2020optimal} and \citet{murali2019noise} utilize SMT solvers for layout synthesis and maximizing circuit reliability or minimizing duration, respectively.

Boolean satisfiability (SAT) solvers represent another pivotal tool in this domain, facilitating the identification of suitable gate placements by determining variable assignments. 
The work of \citet{hattori2018quantum} illustrates the application of SAT solvers in optimizing qubit placement for sub-circuits with adjacent qubits, 
and \citet{meuli2018sat} demonstrate their use in minimizing T and CNOT gates. 
Furthermore, \citet{schneider2023sat} shows the optimal synthesis of Clifford circuits can be solved as an SAT problem.
However, the NP-completeness of SAT problems introduces an exponential computational expense. Efforts by \citet{wille2019mapping} to introduce performance optimizations have achieved near-optimal solutions by navigating the exponentially large search space characterized by $2^{n\cdot m \cdot|G|}$, with $n$, $m$, and $|G|$ denoting the numbers of qubits, physical qubits, and CNOT gates, respectively.

Integer Linear Programming (ILP) emerges as an alternative exact-solver technique, enabling the resolution of problems through integer variables in linear objective functions. For example, a method proposed by \cite{bhattacharjee2019muqut} seeks to reduce quantum circuit depth by addressing a set of nearest neighbor constraints.
\citet{bhattacharjee2017depth}  propose an ILP formulation to minimize the logical depth of quantum circuits while ensuring qubits interact in a nearest-neighbor fashion.

In most cases, solver-based methods can achieve optimal results for small-scale problems. However, these solvers face scalability challenges, as they exhibit poor computational performance when applied to large-scale circuits. Balancing solution quality with runtime efficiency is crucial. To improve computational efficiency, one approach is to design the qubit mapping model to minimize the number of variables. Consequently, a common strategy involves using heuristic functions to reduce runtime to a manageable level. While this may not always yield the optimal solution, it enables the compilation of large-scale quantum circuits.

\subsection{Heuristic-Based Compiler}
\label{subsec:heuristic}
As previously discussed, heuristic-based methods have been comprehensively reviewed in \citet{kusyk2021survey}. Here, we introduce a few representative approaches and recent advancements. A heuristic-based qubit mapping framework is outlined in Algorithm~\ref{trivial_Superconducting_mapping}, where a key challenge lies in determining which gate $g$ to select when no gate in the front layer is executable, as well as in designing an effective $get\_proper\_swap\_path$ function. When routing qubits, an optimal solution would require considering all circuit layers. However, due to computational constraints, researchers typically consider only the first $k$ layers, where $k$ represents the look-ahead depth. A larger $k$ improves solution quality but significantly increases computational cost. To address these challenges, various heuristic methods have been developed to generate hardware-compatible quantum circuits while balancing performance and scalability. These approaches iteratively construct solutions by selecting locally optimal choices at each decision point. While they do not guarantee a globally optimal solution, they provide a practical trade-off between solution quality and compilation efficiency.

\begin{algorithm}[h]
    \caption{Heuristic Mapping Framework}
    \label{trivial_Superconducting_mapping}
    \KwIn{$\mathcal{C}$}
    \KwOut{$\mathcal{C}'$}
    $G \leftarrow get\_dependency\_graph(\mathcal{C})$\;
    Get the initial mapping $\pi$\;
    $\mathcal{C}' \leftarrow \{\}$\;
    \While{$G$ is not empty}{
        \For{$g \in G.front\_layer$}{
            \If{$g$ is executable}{
                $G.remove(g)$\;
                $\mathcal{C}'.append(g)$\;
            }
        }
        Pick a $g$ from $G.front\_layer$\;
        $path \leftarrow get\_proper\_swap\_path(g)$\;
        $\mathcal{C}'.append(path)$\;
        Update mapping $\pi$\;
    }
    \Return $\mathcal{C}'$\;
\end{algorithm}

\textit{Superconducting Device Compiler Framework.}
\citet{zulehner2018efficient} develops efficient method for mapping quantum circuits onto early IBM QX architectures. Their work systematically introduced the challenge of satisfying CNOT constraints while minimizing the number of additional gates required for mapping, laying the foundation for subsequent optimization techniques.

The work by \citet{siraichi2018qubit} and \citet{tannu2019not} stands out in the context of heuristic approaches for NISQ architectures with limited connectivity. Specifically, \citet{siraichi2018qubit} focused on providing local sub-optimal solutions for each CNOT gate, optimizing the gate placement to accommodate the constraints of sparse qubit connectivity. In contrast, \citet{tannu2019not} proposed a heuristic that aims to minimize the total number of quantum gates, thereby addressing the broader challenge of gate overhead in quantum circuits. Targeting special quantum problems, such as reducing decoherence errors, \citet{oddi2018greedy} studied greedy randomized algorithms to optimize the realization of nearest-neighbor-compliant quantum circuits.

SABRE qubit mapper from \cite{li2019tackling} explores the trade-off between reducing the number of quantum gates and quantum circuit depth. Besides, they propose a novel method to generate high-quality initial mapping, which consider the entire circuit. 
Because quantum circuits are inherently reversible, obtain the reverse circuit is natural. All that needs to be done is flip any gates that have parameters and reverse the gate order in the original circuit. Therefore, it's easy for us to use the finial mapping of the origin circuit as the initial mapping to run the reverse circuit. Furthermore, they consider the reverse circuit's final mapping to be the original circuit's initial mapping. Since all gates in the circuit are considered, the updated initial mapping should be better.

\citet{niu2020hardware} propose an approach to enhance the fidelity and execution efficiency of quantum circuits on real quantum hardware. Their algorithm incorporates calibration data into the mapping process, employing a novel heuristic cost function that accounts for both hardware topology and gate error rates. This enables the dynamic selection between SWAP and Bridge gates, optimizing circuit execution.

\citet{zhang2021time} proposes a novel heuristic function on circuit depth and proves its admissibility for the first time, thus one can use an A*-based heuristic research method to find the optimal result. Their experiment result shows that their result is as good as the OLSQ from \cite{tan2020optimal} but with much less compiled time. Their method is far superior to the exact solver because it is scalable up to hundreds of thousands of gates.

\citet{zhu2020dynamic} proposes a dynamic look-ahead heuristic algorithm, which includes an expansion-from-center scheme for initial mapping and a SWAP-based heuristic search algorithm with a novel cost function. The approach dynamically evaluates the impact of SWAP operations, effectively reduces the number of additional gates and runtime.

\citet{park2022fast} presents a fast and scalable qubit-mapping method for NISQ computers, which optimizes quantum circuits to overcome qubit connectivity limitations. The proposed method includes two main stages: an initial mapping using a recursive graph-isomorphism search to quickly generate a scalable initial layout, and a main mapping with an adaptive look-ahead window search to efficiently resolve connectivity constraints by inserting additional gates (SWAP and BRIDGE). It also improves circuit stability by minimizing depth, making it a promising solution for enhancing the efficiency and fault tolerance of quantum computing operations.

\citet{zhu2021iterated} proposes an iterative local search framework. For each iteration, it uses its mapper to find a local optimal solutions and use this local solution to update initial mapping. Furthermore, their framework is permitted to choose the maximum number of iterations, allowing it to trade time for circuit quality. 

\citet{patel2022quest} presents a novel approach to reduce the CNOT gate count in quantum circuits through systematic approximation, aiming to improve output fidelity and mitigate noise in NISQ devices. The proposed method Quest uses circuit partitioning to make the problem scalable, generates low-CNOT-gate-count approximations for each block using approximate synthesis, and combines these approximations to form full circuits.
It also introduces a theoretical upper bound on the process distance to ensure the quality of approximations and uses a dual annealing engine to select ``dissimilar'' approximations that collectively reduce output distance. 

\citet{liu2022not} presents NASSC (Not All Swaps have the Same Cost), an optimization-aware qubit routing algorithm for quantum computing. Unlike traditional methods that focus solely on minimizing the number of SWAP gates, NASSC incorporates a novel cost function that estimates the reduction in CNOT gates due to optimizations like two-qubit block re-synthesis and commutation-based gate cancellation. This approach allows NASSC to make more informed routing decisions, leading to substantial reductions in the number of CNOT gates and circuit depth.

\citet{niu2023enabling} present a Quantum Multi-programming Compiler (QuMC) aimed at enhancing the utilization of NISQ-era superconducting device by enabling the concurrent execution of multiple quantum circuits. The primary goal of QuMC is to mitigate crosstalk errors while running multiple quantum programs simultaneously. 
They introduce a parallelism manager to optimize the number of circuits executed simultaneously, along with two qubit partitioning algorithms: a greedy approach and a heuristic method, designed to allocate reliable qubit regions for each circuit while accounting for crosstalk effects. The study underscores the potential of multi-programming to bridge the gap between current NISQ devices and future large-scale quantum systems.

In \cite{zhu2023variation}, a multi-agent cooperative approach is proposed to tackle the QCC problem. It is important to note that, despite the terminology reminiscent of machine learning, the agents in this study are not driven by learning algorithms. Instead, each agent operates independently, inserting additional gates (such as SWAP gates) based on its local environment state through heuristic functions. Rather than sharing complete solutions, agents exchange partial solutions as shared information, using this iterative process to progressively converge toward an improved success rate.

\citet{fu2024effective} present SWin, a method designed to address the challenge of balancing effectiveness and efficiency in quantum circuit compilation. The authors identify two common patterns in existing greedy mappers—Immediate Execution and Online SWAP Insertion—which frequently result in suboptimal solutions. To overcome these limitations, SWin employs a sliding window approach coupled with an A* search algorithm to iteratively map sub-circuits while maintaining near-optimal results.

\citet{yu2024symmetry} introduces an efficient quantum circuit remapping algorithm that leverages the intrinsic symmetries of quantum processors to address the scalability challenges of quantum circuit compilation. The proposed symmetry-based circuit mapping (SBCM) uses a symmetry-based subgraph matching  method to identify all topologically equivalent circuit mappings by constraining the search space using symmetries. The SBCM algorithm exhibits linear scaling with the number of qubits and is proven to be optimal in terms of time complexity.

\citet{steinberg2024lightcone} introduce an approach to optimizing quantum circuit mapping based on quantum information theory. They propose a lower bound for SWAP gate counts, known as the ``lightcone bound'', which is derived from the quantum Jensen-Shannon divergence between interaction and coupling graphs. The method optimizes qubit assignments and iteratively minimizes divergence through the use of Bell measurements and doubly-stochastic quantum channels.

\citet{escofet2024route} introduces a novel quantum circuit mapping algorithm designed to address the challenges of scalable quantum computing. The algorithm incorporates attraction forces between qubits to guide routing, prioritizes high-fidelity links in modular architectures, and includes tunable hyperparameters to balance trade-offs between compilation time, circuit depth, and gate overhead.

\citet{cheng2024robust} proposes a novel routing algorithm called Duostra (Dual-source Dijkstra) to solve the quantum circuit compilation problem of large-scale circuits. Unlike previous studies which treat the qubit mapping problem as a whole, they divide the problem into two sub problems - routing and scheduling. For each two-qubit gate, the router will use Dijkstra algorithm to find a routing path that minimizes the execution time. The scheduler only needs to consider the order of the gates for execution.

Overall, while the heuristic framework may not yield an optimal solution, it offers significant flexibility in balancing mapping cost and compilation runtime for large-scale circuits. Additionally, the results indicate that as circuit size increases, the significance of initial mapping diminishes, while routing becomes increasingly critical.

\subsection{AI-Based Compiler}
\label{subsec: ai sc}

The application of AI-based methods to \textbf{M\&S} problem presents a promising approach to overcoming the challenges of optimizing quantum circuits on real hardware. AI techniques offer flexibility and adaptability by enabling algorithms to learn from past experiences and hardware-specific constraints. These AI-driven methods can dynamically adjust to variations in hardware performance, resulting in robust and efficient solutions. A survey of various \textit{AI planning} algorithms, including Planning Domain Definition Language (PDDL) and the Stanford Research Institute Problem Solver is provided by \citet{kusyk2021survey}. In this context, we focus on the recent advancements in methods using \textit{machine learning} and \textit{reinforcement learning} for improving quantum circuit optimization.

\textbf{Reinforcement Learning. }
\citet{huang2022reinforcement} proposed a model based on reinforcement learning to solve the initial mapping problem and also proposed a DEAR (Dynamic Extraction and Routing) framework to solve the SWAP insertion problem based on A* search. They build an encoder-decoder model as the agent and set an existing qubit routing tool as the environment. Reinforcement learning can help maximize the reward and get a good initial mapping. The experimental results show that their model can lead to 12\% fewer additional gates. Furthermore, \citet{pascoal2024deep} propose using deep reinforcement learning to address the problem, employing Proximal Policy Optimization (PPO) as the training algorithm.

\textbf{Machine Learning. }
\citet{sinha2022qubit} proposed an architecture enhanced by Graph Neural Networks (GNN) \cite{wang2019dynamic} to optimize quantum qubit routing. They employ Monte Carlo Tree Search (MCTS) to perform the routing task, with the GNN utilized to evaluate the reward and update the MCTS accordingly, thereby improving the overall efficiency and accuracy of the routing process. \citet{paler2023machine} introduce the heuristic QXX and its machine learning-based counterpart QXX-MLP. QXX uses a Gaussian-like function to estimate the depth of the laid-out circuit, while QXX-MLP approximates the circuit depth more efficiently. \citet{ren2024leveraging} introduce a Transformer-based model trained to predict SWAP gate insertions required to satisfy hardware connectivity constraints. Unlike traditional heuristic or search-based methods, the model takes as input a window of the quantum circuit—represented as a gate sequence with hardware-aware encoding—and outputs a ranked list of candidate SWAPs to apply before the next gate layer. 
The output of the model will be evaluated by the evaluate function. If it is deemed sufficiently good, the swap sequence generated by the model will be adopted and applied to update the system. Otherwise, a ``fallback'' heuristic will be employed to resolve the first non-executable gate.
They also introduces several enhancements, including input representations that encode both gate type and topological context, and training strategies that align with realistic compilation objectives.

\subsection{Application-Specific Compilers}\label{subsec:application_specific}
Beyond general-purpose quantum compilers, specialized compilers leverage application-specific program patterns for enhanced performance. Some representative real-world applications or program kernels are promising candidates to showcase practical quantum advantages, driving extensive research into application-specific compilers, particularly in superconducting devices.

\textbf{Hamiltonian Simulation. } Variational quantum algorithms (VQA) based on Hamiltonian simulation represent a specialized class of quantum programs (e.g., VQE for chemistry simulation and condensed-matter physical system simulation~\cite{peruzzo2014variational}, QAOA for combinatorial optimization~\cite{farhi2014qaoa}) well-suited for near-term quantum computing applications due to its modest resource requirements in terms of qubits and circuit depth~\cite{feynman2018simulating}. Since Hamiltonian simulation programs are essentially composed of disciplined subroutines known as Pauli exponentials (Pauli strings with coefficients) that are variably arranged, various compiler schemes have been proposed to harness these distinct characteristics to perform extreme compilation optimization. For the general Hamiltonian simulation programs, representative dedicated compilation approaches include ZX diagrams of Pauli gadgets (e.g., TKet~\cite{cowtan2019phase}, PCOAST~\cite{paykin2023pcoast}, PauliOpt~\cite{meijer2023towards}), synthesis variants of Pauli-based IR (e.g., Paulihedral~\cite{li2021paulihedral}, Tetris~\cite{jin2024tetris}), and IR simplification based on BSF simplification (e.g., Rustiq~\cite{de2024faster}, Phoenix~\cite{yang2025phoenix}). For the special 2-local Hamiltonian simulation programs (e.g., QAOA, Heisenberg model), some special compilers like 2QAN~\cite{lao20222qan} tend to exploit more optimization opportunities and achieve better results than the general VQA compilers above. In the limited-connectivity NISQ scenario, soft-hardware co-optimization VQA compilation strategies could not only lead to excellent circuit synthesis effects but also guide the efficient NISQ processors design for Hamiltonian simulation programs. For example, \citet{lao20222qan} introduced 2QAN, a quantum compiler tailored for 2-local Hamiltonians, featuring the ``unitary unifying'' technique, which reduces gate count by merging consecutive two-qubit gates. \citet{li2021paulihedral} proposed Paulihedral, which optimizes Pauli term order to minimize circuit depth and adapts each term for limited-connectivity in NISQ superconducting devices. Additionally, \citet{alam2020efficient, alam2020circuit, alam2020noise} introduced ZZ, focusing on the commutation of ZZ gates in QAOA~\cite{farhi2014qaoa}. \citet{li2021software} explore a software-hardware co-design framework through the proposed $X$-tree architecture, which also optimizes performance by pruning Pauli strings based on their significance.

\textbf{Quantum Machine Learning (QML).} Many practical problems in quantum computing can be framed as ground state preparation, where parameterized quantum circuits or quantum neural networks offer a natural and powerful solution. However, factors such as noise, expressibility, and trainability significantly influence the final performance. To balance these trade-offs, it is essential to automate the design of quantum circuits—a process known as Quantum Architecture Search (QAS). QuantumSupernet~\cite{du2022quantum} and QuantumNAS~\cite{wang2022quantumNAS} are built upon the classical Supernet framework~\cite{pham2018efficient}, but they rely heavily on computationally expensive gradient calculations. In contrast, the work presented in~\cite{anagolum2024elivagar} introduces representational capacity as a metric to measure intra-class similarity and inter-class separation, providing an efficient method for reducing the training overhead in Supercircuit-based QML tasks.
In addition to focusing on the training performance of QML, some QAS methods also pay attention to the impact of noise and strive to mitigate its influence on QML tasks (e.g., QuantumNAT~\cite{wang2022quantumnat}, QuantumNAS~\cite{wang2022quantumNAS}).
However, these QAS methods are training-based, which incurs substantial resource overhead. To address this challenge, many works attempt to explore more scalable QAS approaches. For instance, \citet{he2024training} proposed a method that randomly generates quantum circuits and then selects the optimal ones, thereby avoiding the training overhead. Meanwhile, \citet{zhu2025scalable} introduced an approach that leverages Clifford sampling to bypass the computational overhead associated with training.
In addition, \citet{quetschlich2024towards} presented a case study on quantum generative models, demonstrating that QML techniques can be effectively applied to the task of quantum circuit compilation. Specifically, the authors highlighted the use of AI-driven methods to generate Quantum Circuit Born Machines (QCBMs), which are subsequently employed to produce optimized compiled quantum circuits.

\textbf{Quantum Error Correction (QEC). } Quantum computing has made significant progress with the development of NISQ devices across various hardware platforms, enabling early applications such as variational quantum algorithms. However, due to inherent noise and limited scale, quantum error correction is widely recognized as the essential path toward fault-tolerant quantum computing, which is necessary for solving large-scale, practical problems.

\textit{Lattice Surgery Compilers.} Among QEC codes, the surface code is a leading candidate due to its high noise threshold ($\sim$1\%~\cite{google2023suppressing}), topological layout, and compatibility with universal quantum computation via magic-state distillation. Techniques like lattice surgery enable scalable multi-qubit operations, but efficient algorithm execution still requires careful resource management. In the line of logical qubit mapping and lattice surgery scheduling, several works~\cite{lao2018mapping, molavi2025dependency} propose quadratic assignment and SAT-based formulations to optimize them. However, given that lattice surgery optimization is NP-hard~\cite{herr2017optimization}, scalable methods are essential. \citet{litinski2019game} formulate the problem as a game and propose heuristic functions to analyze the trade-off between volume and time costs in the proposed layout. Although \citet{watkins2024high, leblond2023realistic} propose general frameworks for translating circuits into lattice surgery operations, they overlook opportunities to minimize the underlying scheduling costs. LaSsynth~\cite{Tan2024LaS} proposes a SAT-based solver that can optimally handle a limited number of qubits and operations, whereas \citet{liu2023substrate} introduces a substrate scheduler capable of efficiently compiling graph states with thousands of vertices. Since the implementation of two-qubit gates requires occupying a physical path on the layout, and simultaneous paths cannot intersect, gates that could theoretically be executed in parallel may instead be forced into sequential execution. To address this limitation, several works~\cite{beverland2022assessing, beverland2022surface, hamada2024latticesurgery, hirano2025localityaware} have proposed techniques to enhance circuit parallelism in lattice surgery.

In the other surface code lattice surgery pipeline optimization, there are also many works focusing on it to further lower the overall cost of executing algorithms. For example, ~\cite{wang2024optimizing} introduced TACO to reduce the Clifford cost, primarily focusing on optimizing the Clifford+T decomposition by minimizing the number of $R_Z$ gates and~\cite{chatterjee2025qspellbook} introduced Q-Spellbook to help users select the suitable data block layouts and distillation protocols under different optimization strategies. \cite{ueno2024high} proposes a Bypass architecture composed of dense and sparse qubit layers, enabling multiple lattice surgery path options and improving the efficiency of long-path lattice surgery operations using fewer qubits. \cite{kan2025sparo} introduces SPARO, which actively identifies bottlenecks caused by constraints such as limited routing areas or magic-state factory throughput, and dynamically allocates hardware resources to balance or mitigate them.

Beyond the surface code, other topological codes such as the color code and folded surface code also support planar layouts with geometrically local but not strictly nearest neighbor connectivity, potentially reducing qubit overhead. \cite{herzog2025lattice} present a general compilation framework based on the concept of a code substrate, enabling scalable lattice surgery compilation for these codes through logical qubit allocation and efficient routing strategies. There are many other works have also explored heterogeneous designs for FTQCs. For example, \cite{stein2023hetarch} proposes a toolbox for exploring heterogeneous FTQC architectures, demonstrated with superconducting devices, while \cite{xu2024constant, bravyi2024high, stein2025hetec} propose combining different QEC codes, such as surface codes and qLDPC codes, to leverage the advantages of both.

\textit{QEC code implementation.} Many efforts have focused on improving the efficiency of fault-tolerant quantum computation through compilation methods and these approaches often assume that logical qubits or existing quantum error correction codes can be efficiently implemented on current quantum hardware. However, implementing QEC codes remains complex in practice due to the mismatch between code topology and hardware architecture. For example, surface codes require a two-dimensional lattice of qubits, but most superconducting devices do not yet support this structure. Towards addressing these problems, several studies have proposed manually designed architectures tailored to specific quantum error-correcting codes. For example, planar qubit layouts have been introduced for synthesizing color codes~\cite{reichardt2020fault}, while trivalent architectures have been developed for implementing triangular color codes~\cite{chamberland2020triangular}. In order to avoid manually redesigning code protocols, \cite{wu2022synthesis} proposes Surf-Stitch, an automatic synthesis framework for implementing and optimizing surface codes on superconducting devices. Surf-Stitch uses a three-stage modular optimization pipeline tailored to superconducting hardware constraints. It first performs data qubit allocation by selecting compact, conflict-free regions using high-degree qubits. Then, it constructs efficient bridge trees for measurement circuits using star-tree or branching-tree strategies to minimize depth and maintain locality. Finally, a heuristic-based measurement scheduling algorithm maximizes parallelism while avoiding conflicts. \cite{yin2025qecc} further extend this line of work by focusing on mapping stabilizer codes to hardware using an ancilla bridge technique~\cite{chamberland2018flag,chao2018fault,lao2020fault} and a MaxSAT-based optimization framework. Regarding other implementation challenges in superconducting devices, \cite{yin2024surf} presents a code deformation framework with adaptive defect mitigation tailored to specific defect scenarios, while \cite{fang2024caliscalpel} proposes CaliScalpel for in situ calibration in surface code operations.

\section{Trapped-Ion Compiler}
\label{sec:trapped-ion}
\subsection{Background}
Trapped-Ion (TI) have been a leading technology platform for quantum computing. Key milestones include the achievement of single-qubit gates, two-qubit gates, and qubit state preparation and readout with high fidelity~\cite{levine2018high,gaebler2016high,crain2019high}, exceeding requirements for fault-tolerant quantum computing using high-threshold quantum error correction codes. The development of TI quantum computing, as a pivotal branch of quantum information science, exemplifies the transition of quantum technologies from theoretical constructs to practical applications. 

The genesis of TI quantum computing can be traced back to the 1990s, marked by seminal contributions from Nobel laureates such as David Wineland~\cite{leibfried2003quantum}, along with other scientists like Ignacio Cirac, Peter Zoller~\cite{cirac1995quantum}. The foundational proposal by Cirac and Zoller in 1995 to use individual atomic ions confined in radiofrequency (RF) traps as qubits marked a critical moment in this field, and later a more feasible proposal given by M\o lmer and S\o rensen which do not require the ion cooled to the ground motion state, significantly improving the practicability of trapped-ions~\cite{sorensen1999quantum}. In recent years, advancements in trap design and electrical potential control have led to improvements in fast shuttling, ion separation, optical phase control, and junction transport, crucial for enhancing the efficiency of trapped-ion quantum computers~\cite{bruzewicz2019trapped}.

Trapped-Ion quantum computing platforms offer several distinctive advantages. Firstly, they enable high-fidelity quantum gate operations, crucial for the accurate execution of quantum algorithms. Secondly, TI devices exhibit prolonged coherence times for quantum states~\cite{wang2021single}, facilitating the execution of more complex quantum algorithms. Lastly, a unique feature of TI is their fully connected qubits, allowing for direct interaction between any two qubits, simplifying quantum circuit design.

\begin{table}[t]
\caption{A summary of trapped-ion compilers in literature.}
\resizebox{1.0\linewidth}{!}{
\renewcommand{\arraystretch}{1.5}
\begin{tabular}{c|c|c|c|c|c|c }
\toprule

\textbf{Ref.}& \textbf{Year} & \textbf{Architecture} &\textbf{Problem} &  \textbf{Algorithm} &   \textbf{AOM} & \textbf{Physical Swap}   \\ \midrule
\cite{wu2019ilp_iontrap}& 2019& Linear Type &\textbf{M\&S} &Integer Linear Program &   16& - \\ 
~\cite{murali2020architecting_iontrap}& 2020  & QCCD   &\textbf{M\&S} & Heuristic   & All  & \checkmark  \\ 
~\cite{wu2021tilt_iontrap}& 2021 & Linear Type&\textbf{M\&S} &   Heuristic     & 16/32    & -   \\ 
~\cite{saki2022muzzle_iontrap}& 2022& QCCD  &\textbf{M\&S}&Heuristic    & All& - \\ 
~\cite{upadhyay2022_iontrap}& 2022 & QCCD & \textbf{M\&S}&Heuristic   & All& -  \\ 
~\cite{schmale2022backend_iontrap}& 2022  & QVLS-Q1 chip (QCCD)&\textbf{M\&S} &  Heuristic  & -   & -\\ 
~\cite{durandau2022automated_iontrap}& 2022  & Linear Segmented (QCCD)&\textbf{M\&S} &  Heuristic   & 1 & \checkmark  \\ 
~\cite{leblond2023tiscc}& 2023 & 2D square grid (QCCD)& QEC Compiler &  Code Mapping  & - & -  \\ 
~\cite{kreppel2023quantum_iontrap}& 2023 & Linear Segmented (QCCD)&\textbf{M\&S} &  Heuristic   & 1 & \checkmark  \\ 
~\cite{dai2024advanced}& 2024 & QCCD &\textbf{M\&S}& Heuristic  & All& -  \\ 
~\cite{Chu2024titan}& 2024 &  Distributed TI computer&\textbf{M\&S} & Heuristic  & All & - \\ 
\cite{schoenberger2024using}& 2024& 2D square grid (QCCD)&\textbf{M\&S} &  SAT Solver & All & -  \\ 
\cite{ schoenberger2024shuttling}& 2024& 2D square grid (QCCD) &\textbf{M\&S}&  Heuristic & All & -  \\
\cite{preti2024hybrid}& 2024& Trapped-Ion Circuits& Circuit Synthesis &    Reinforcement Learning & - & -   \\ 

~\cite{wu2025boss}& 2025 & Linear Type & \textbf{M\&S}  &Heuristic     & 16/32    & -   \\ 
\cite{zhu2025s}& 2025& QCCD & \textbf{M\&S} &    Heuristic  & All & -   
\\ 
\cite{bach2025efficient}& 2025& QCCD & \textbf{M\&S} &    Heuristic  & All & -   
\\ 
\cite{ruan2025trapsimdsimdawarecompileroptimization}& 2025&  2D square grid (QCCD)    & \textbf{M\&S} &    Heuristic  & All & -   
\\\cite{yin2025flexion}& 2025&  2D square grid (QCCD)    & QEC Compiler &   Heuristic    & All & -   
\\
\bottomrule
\end{tabular}
}
\label{tab:summarizations_iontrap}
\end{table}

However, alongside these advantages, scalability presents a significant challenge. Physical constraints such as the stability and control complexities of traps escalate with increasing qubit numbers. Additionally, the proliferation of control lasers for larger qubit arrays can lead to cross-talk and interference, posing further operational challenges. 
In response to these scalability challenges, researchers have been exploring novel approaches~\cite{kielpinski2002architecture}. 
The quantum charge-coupled device (QCCD) architecture represents a promising solution, partitioning traps to enable localized control over smaller ion groups, thus enhancing operability in large-scale systems. Another innovative approach is the shuttle scheme, which involves moving ions within microfabricated TI arrays. This method facilitates the efficient transfer of quantum information, enabling the transmission of quantum states between different processing units. These approaches not only address the immediate scalability challenges but also open new avenues for the future development of TI quantum computing. This review explores compiler design in the evolution of TI devices, tracing their progression from \textbf{linear tape devices} to \textbf{quantum charge-coupled devices} and ultimately to \textbf{large-scale distributed systems}. 
We summarize the relevant compiler for the trapped-ion devices, and the content is summarized in Table \ref{tab:summarizations_iontrap}.

\subsection{Linear Tape Device}
\label{subsec:linear tape}
The single trap trapped-ion device, specifically the Trapped-Ion Linear-Tape (TILT), represents one of the earliest and most fundamental trapped-ion architectures. TILT is a linear, Turing-machine-like structure that employs a multi-laser control ``head'', where a linear chain of ions moves back and forth under the laser head \cite{wu2019ilp_iontrap}. This technology is well-established and offers several advantages. Firstly, its simplicity is a key strength. TILT avoids the complex components required for scaling up other architectures, thereby reducing technical barriers. Secondly, TILT enables parallel gate operations on multiple qubits, which significantly enhances computational efficiency compared to the sequential approach of QCCD. Lastly, although TILT is not inherently scalable, the ion scheduling within a single trap provides insights that are analogous to the processes in multi-trap trapped-ion devices. This offers valuable perspectives on ion management, offering insights that are highly applicable to other types of trapped-ion devices.

\begin{figure}[H]
    \centering
    \includegraphics[width=0.55\linewidth]{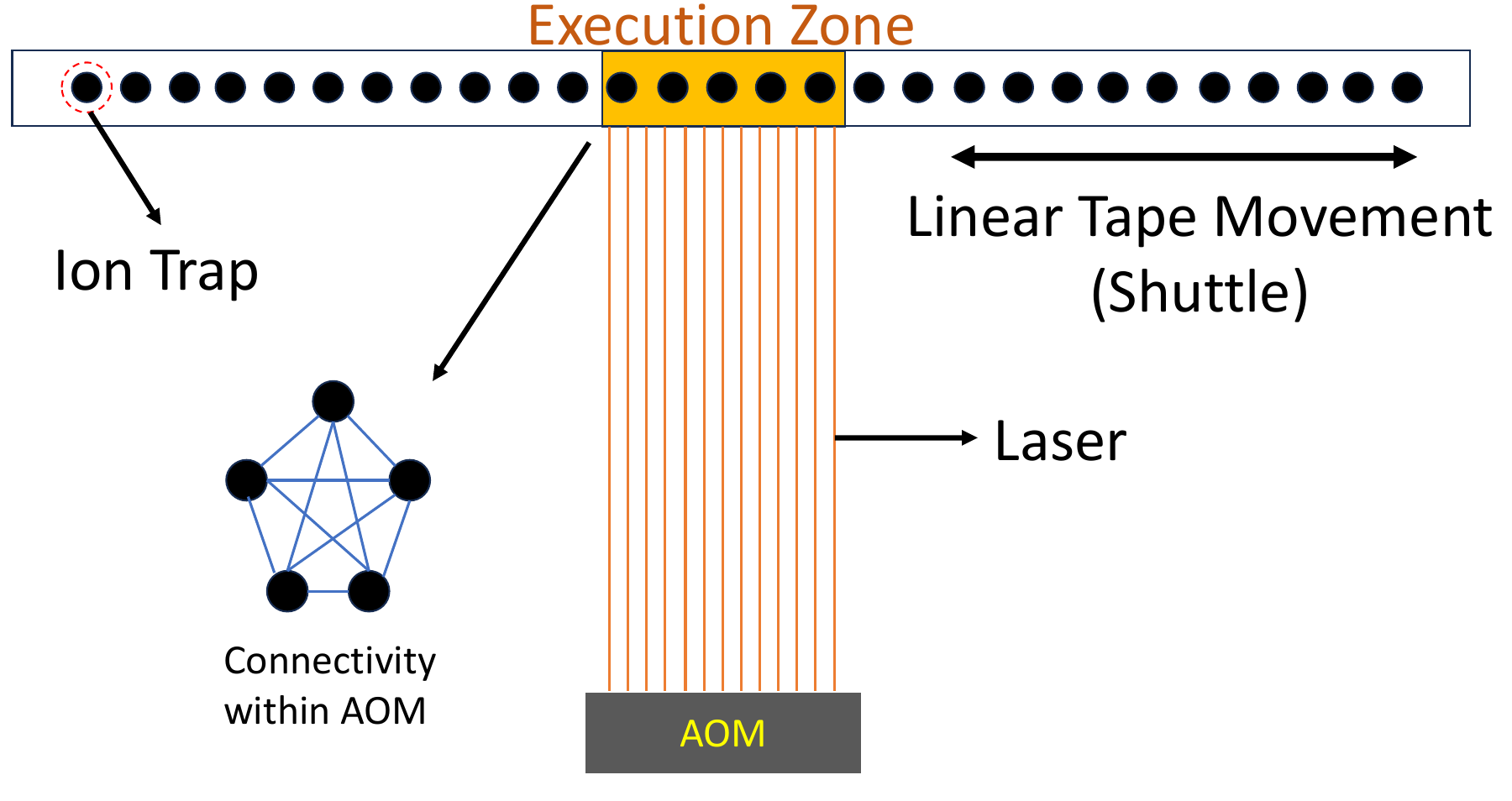}
    \caption{An illustration of linear-tape trapped-ion chips demonstrates that qubits within the Acousto-Optic Modulator (AOM) zone are fully interconnected. To execute operations involving qubits outside the AOM zone, ion shuttling becomes necessary.}
    \label{fig:tilt}
\end{figure}
\textbf{Constraints. }An illustration of linear-tape trapped-ion devices can be found in Fig.~\ref{fig:tilt}. In a TILT system, full connection between all qubits is no longer provided because the tape head is not covering the entire ion chain. It is only fully connected within the execution zone. A two-qubit gate is executable if the distance between the two qubits involved is equal to or less than the size of the tape head. There is no need for more gates, although it could be necessary to relocate the entire chain behind the head. Nevertheless, both swap gates and tape motions are required if the distance between the two qubits is greater than the size of the tape head. Moreover, no gate—not even a single-qubit gate—may be used outside of the execution zone; all gates must operate within it.

Superconducting compiler methods are not directly suited for trapped-ion systems, as they assume a static hardware graph where two-qubit gates are constrained by fixed edges, with logical qubits remapped using SWAP gates. In contrast, the TILT architecture features a dynamic topology---a fully connected subgraph where gate execution is restricted and other isolated vertices. Unlike superconducting systems, these isolated vertices cannot be remapped via SWAP gates, requiring a different compilation approach. This modification increases the complexity of quantum algorithm implementation within the TILT framework, necessitating the compiler to accommodate the graph's dynamic behavior.

\textbf{Problem Definition.} In linear-tape trapped-ion chips, the qubit mapping problem can be defined as follows: given the quantum circuits, the AOM size and noise conditions, determine the mapping between logical and physical qubits, along with the linear tape shuttle schedules, to ensure that the mapped circuit can be executed within the AOM zone while minimizing its impact on circuit execution results.

\textbf{Noise Simulator.} Trapped-ion devices are generally less accessible than superconducting devices. A common approach for evaluating the performance of scheduling algorithms is through noise simulation. The gate fidelity is typically modeled after $m$ moves as~\cite{wu2021tilt_iontrap},
\begin{equation}
    F_m = 1- \Gamma \tau + (1 - (1+\epsilon)^{2mk+1})
\end{equation}
where $\Gamma$ is the background heating rate of the trap, $\tau$ is the gate time as a function of number of intraion spacings between the involved qubits, $k$ is the average amount of heating added to the chain during each shuttle and $\epsilon$ is the error in each two-qubit gate due to residual entanglement with the motion.

The accumulation of heat during shuttling operations is a primary cause of fidelity degradation. This heat-induced decline can be effectively mitigated through the application of cooling lasers, which can eliminate the excess thermal energy introduced by shuttling. 
In addition, it is important to note that the act of shuttling itself remains an inherent source of fidelity loss, even with the implementation of cooling techniques. The fidelity loss due to the \( m \)-th shuttle operation can be estimated by,
\begin{equation}
F_{\text{shuttle}} = 1 - \epsilon_{\text{shuttle}} m,    
\end{equation}
where \( \epsilon_{\text{shuttle}} \) is the per-shuttle fidelity loss, and \( m \) is the number of shuttles \cite{wu2025boss}.

Implementing a two-qubit gate is also challenging. The ion-ion coupling strength for a pair of ions at distance $d$ scales in proportion to $1/d^\alpha$ with $\alpha$ ranging from 1 to 3~\cite{zhang2017observation, leung2018entangling}, which means the requirement of a longer pulse time to entangle more distant ions. Furthermore, the collective motional modes (phonons) of the ion chain are used to mediate the two-qubit interaction. The density of modes increases with ion number, worsening the chance of crosstalk among modes, hampering the ion chains cooling and initial states preparation, all these reducing the gate fidelity in trapped-ions quantum computing.

\textbf{Linear Tape Compilers. } Similar to superconducting chips, the proposed method for addressing the mapping and routing problem in linear-tape trapped-ion chips involves both \textbf{solver-based} and \textbf{heuristic} approaches. 

\textit{TILT compilation framework.} We first introduce a general compilation framework for TILT devices that incorporates the concept of superconducting devices, as shown in Algorithm~\ref{trivial TILT mapping}. The functions $FindProperSwap()$ and $FindProperTapeMovement()$ are two crucial components, which are primarily implemented using heuristic functions. These heuristic functions can be based on existing research on superconducting devices to enhance their effectiveness and applicability.

\begin{algorithm}[h]
    \caption{TILT Mapping Framework}
    \label{trivial TILT mapping}
    % \SetAlgoLined
    \KwIn{$\mathcal{C}$}
    \KwOut{$\mathcal{C}'$}
    $\mathcal{C}' \leftarrow \{\}$\;
    Get initial map $\pi$ and initial execution zone $Z$\;
    \While{true}{
        $g \leftarrow get\_next\_gate(\mathcal{C})$\;
        \If{$g$ is executable}{
            $\mathcal{C}' \leftarrow \mathcal{C}' \cup \{g\}$\;
            update $\mathcal{C}$\;
        }
        \ElseIf{$g.q_1$ or $g.q_2$ is in $Z$}{
            $swap \leftarrow FindProperSwap()$\;
            update mapping $\pi$\;
            $\mathcal{C}' \leftarrow \mathcal{C}' \cup \{swap\}$\;
        }
        \Else{
            $tape\_move \leftarrow FindProperTapeMovement()$\;
            update execution zone $Z$\;
            $\mathcal{C}' \leftarrow \mathcal{C}' \cup \{tape\_move\}$\;
        }
    }
    \Return $\mathcal{C}'$\;
\end{algorithm}

\citet{wu2019ilp_iontrap} introduced STRIQC, a solver-based method that employs Integer Linear Programming (ILP) to optimize the scheduling of quantum gates on linear tape devices. By using \( c_{t,g} \) to indicate whether gate \( g \) is executed in time slot \( t \), the ILP model ensures optimal results for quantum circuit execution. 
In contrast, \citet{wu2021tilt_iontrap} focused on enhancing the utilization of TILT devices through two heuristic functions. The first function optimizes tape movement by selecting the tape head position with the maximum number of executable gates, while the second employs the Opposing Swap technique to minimize the number of inserted SWAP gates, ensuring that all two-qubit gates are within a predefined distance threshold before scheduling. 
More recently, \citet{wu2025boss} proposed BOSS, a novel compiler for TILT architecture that addresses the challenge of maintaining high gate fidelity by minimizing shuttling operations. BOSS uses a blocking algorithm to cluster gates together, reducing qubit idle rates in the execution zone, and then employs an efficient scheduling algorithm to orchestrate these blocks. Additionally, BOSS incorporates cooling technology \cite{chen2020efficient} to further mitigate losses from shuttling operations.

\subsection{Quantum Charge-Coupled Devices}
\label{subsec:qccd}
\textbf{Background.} Single trap devices, although foundational to the evolution of trapped-ion technologies, encounter several challenges that impede their scalability. For instance, with an increase in the number of ions, the density of vibrational modes escalates, which in turn heightens the probability of crosstalk between these modes. This crosstalk can lead to a significant decrease in the fidelity of quantum gates. In response to these limitations, researchers have developed multi-trap devices, which offer enhanced scalability and mitigate some of the issues associated with single trap architectures.
Multi-trap trapped-ion devices are generally configured as Quantum Charge-Coupled Devices (QCCD). This architecture allows for the physical shuttling of ions between different traps, facilitating qubit transfer and operations. The QCCD setup enables dynamic movement and manipulation of ions across various processing regions, making it a promising approach for scaling trapped-ion qubits.

\begin{figure}[h]
    \centering
    \includegraphics[width=1.0\linewidth]{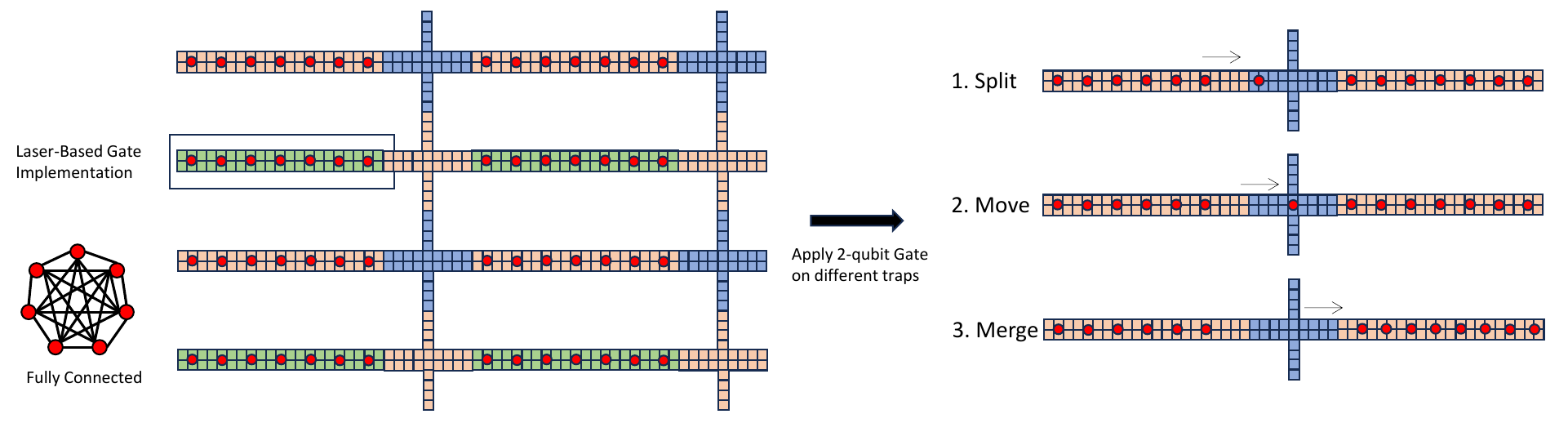}
    \caption{A modular Quantum Charge Coupled Device (QCCD) consists of several traps, each of which is initialized with 7 qubits. These traps are interconnected with shuttle paths. In order to enable the implementation of a two-qubit gate between two traps, ions need to be split from one trap, moved, and then merged into another trap by traversing the shuttle path.}
    \label{fig:qccd_whole}
\end{figure}

\textbf{Constraints.}
Given the modular nature of QCCD, ensuring connectivity between ions across different modules is of utmost importance. The critical operations involved are ``split'', ``move'', and ``merge'', as illustrated in Fig.~\ref{fig:qccd_whole}. 
Specifically, the ``split'' operation involves spatially separating a single ion from an ion crystal, effectively isolating it from interactions with other ions. In contrast, the ``merge'' operation is the reverse process, where an isolated ion is reintegrated into an ion crystal, thereby becoming part of the chain and enabling participation in two-qubit gate operations with other ions.

\begin{figure}[H]
    \centering
    \includegraphics[width=0.5\linewidth]{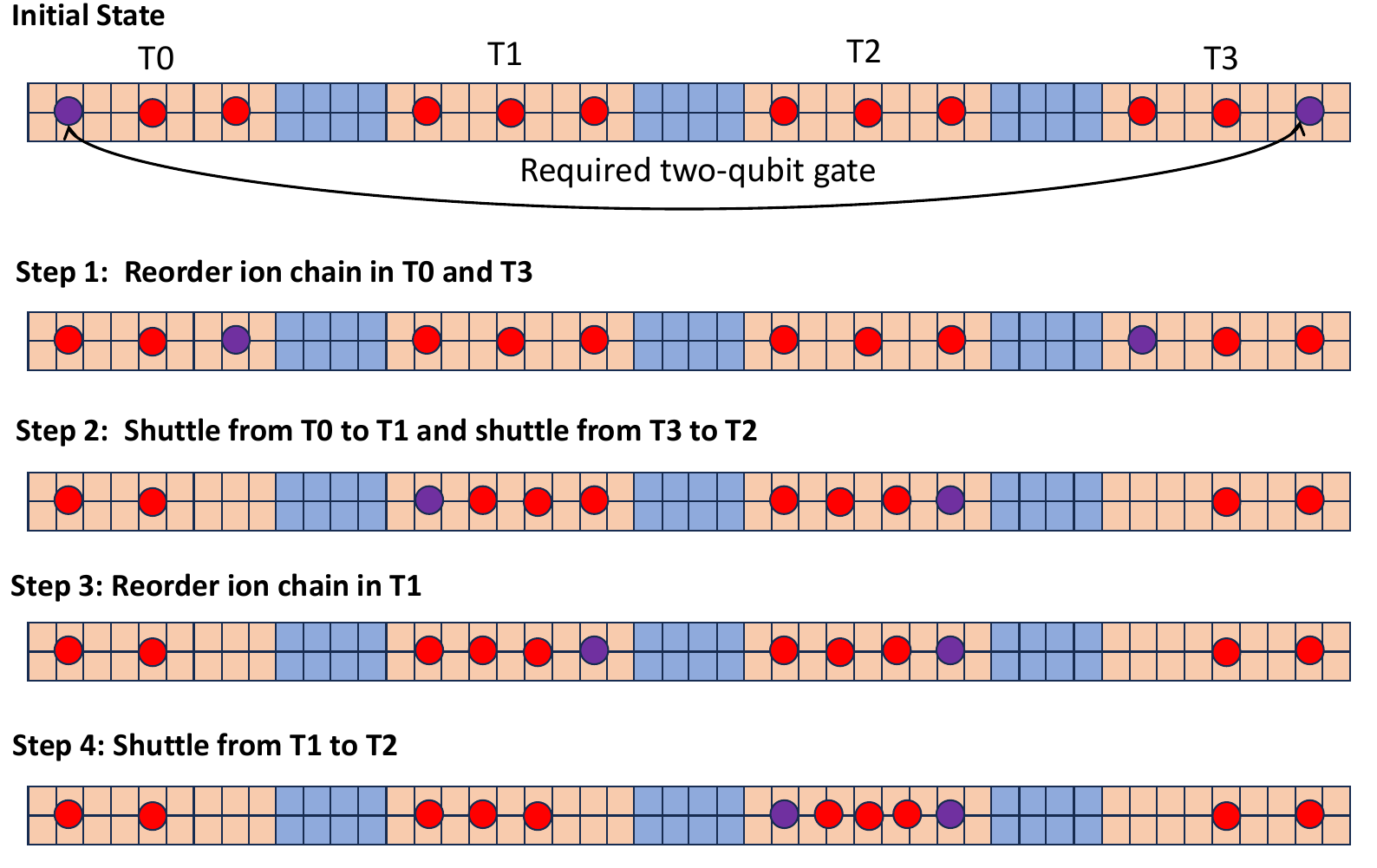}
    \caption{ Shuttling in a QCCD system which has linear device topology. Extra split and merge operations are required while moving ions through intermediate traps.}
    \label{fig:LNN_QCCD}
\end{figure}
QCCD systems employ ion shuttling, which differs slightly from the linear type, to connect traps. Here, ``shuttle'' refers to the act of physically moving qubits to enable inter-trap communication. An example is shown in Fig.~\ref{fig:LNN_QCCD}, which is a linear QCCD without any junctions. For each shuttle, it has 1 split, 1 move and 1 merge. Each shuttling path can only have one ion used, but they can proceed simultaneously if these shuttles act on different segment.

In QCCD systems, the topology is dynamically altered during operations. The shuttle operation modifies the topology graph, indicating that the physical connections between qubits can change during the process. This dynamic nature allows for flexible qubit interactions but may introduce additional complexity and potential error sources.

\textbf{Noise Simulator.} The operations of separation, merging, swapping, and shuttling, required for executing two-qubit gates between different modules, involve the movement of ions. While the quantum information stored in the ions' internal states remains unchanged during spatial movement, the motional mode (phonon) used to connect the ions is strongly affected by these movements.
These operations involve the acceleration and deceleration of ions through micro-electrode control, which can significantly heat the ion chain and impact the precision of subsequent two-qubit gates. Moreover, ion qubit movement operations incur a time cost. If the coherence time of the ions' internal states is short, the additional time required for these operations can reduce the overall fidelity of the quantum circuit and operations.

Considering the energy of an ion chain is predominantly constituted by the kinetic energy of individual ions and the Coulomb potential energy between ions, the process of splitting and then merging ion chains leaves the overall Coulomb potential energy unaffected. Nevertheless, the total kinetic energy of the ion chain experiences an increment of $k_1$ quanta of motional energy, a consequence of the manipulation imposed by external electrodes. When the qubit is being shuttled from segments, it adds $k_2$ quanta of energy. The combined effect of $k_1$ and $k_2$ leads to an increase in the total occupation number of phonon modes in the ion chain, which is typically denoted as $\bar{n}$. The product $A \bar{n}$ quantifies the overall transport impact on fidelity, where $A\propto N/\ln(N)$ is a scaling factor which represents the thermal laser beam instabilities. As the total number of ions $N$ in the chain changes with each shuttle operation, the product $A \bar{n}$ will change with every transport operation.

Therefore, we can establish a correlation between the fidelity of two-qubit gates in trapped-ions and factors such as the number of ions in the chain, the operation time $\tau$, and the background heating efficiency of the trap electrodes $\Gamma$. The model can treat the background heating rate as a constant. The total heat accumulation is given by $\Gamma \tau $ over different operations. Using above analysis model, the cumulative impact of these factors on the fidelity of the quantum circuit can then be represented by,
\begin{equation}\label{eq:fidelity}
    F = 1 - \Gamma\tau - A(2\bar{n} + 1).
\end{equation}

\textbf{QCCD Compilers. }
Subsequently, we will delve into an exploration of other QCCD compilers.

\citet{murali2020architecting_iontrap} first introduce a compiler that optimizes qubit operations. Firstly, it maps program qubits to distinct hardware qubits using heuristic techniques to minimize communication. Then, it addresses qubit shuttling by routing through the hardware's shortest pathways and inserting necessary chain reordering operations. On QCCD devices, the compiler efficiently routes parallel shuttles to prevent deadlocks and congestion, enhancing system performance.

\citet{murali2020architecting_iontrap} also examine the impact of micro-architecture design on performance. Their study shows that increasing trap capacity from 20 to 35 ions results in a 3$\times$ increase in error rate. Additionally, they identify an optimal trap capacity of 15–25 ions, depending on the application, which minimizes heating effects from communication, laser beam thermal motion, and localized hot spots. This trap sizing also provides strong runtime performance across various applications.
Furthermore, they show that topology plays a significant role in determining the fidelity of SquareRoot and QFT circuits. Specifically, for SquareRoot, the grid topology achieves up to 7000× higher fidelity compared to the linear topology. In contrast, for QFT, the linear topology outperforms the grid topology, achieving up to 4$\times$ higher fidelity.

\citet{saki2022muzzle_iontrap} introduce three heuristic approaches for optimizing shuttling strategies in \textit{linear-connected QCCD devices}: future operations-based shuttle direction policy, nearest-neighbor-first rebalancing with improved ion selection, and opportunistic gate reordering. Future operations-based shuttle direction policy dynamically adjusts shuttle movement to enable two-qubit gate execution, with a time complexity of $O(n^2)$. When incorporating gate proximity, the complexity is reduced to $O(nk)$, where $k \leq n$. Opportunistic gate reordering determines a shuttle direction for each pending gate and evaluates them across the active layer and previous layers, maintaining a complexity of $O(n^2)$. Nearest-neighbor-first rebalancing enhances ion reallocation by prioritizing the closest available traps, achieving a time complexity of $O(n)$. Experimental results demonstrate a 19\% to 51\% reduction in the number of shuttling operations, significantly improving efficiency.

The initial mapping is less explored in~\cite{murali2020architecting_iontrap, saki2022muzzle_iontrap}. \citet{upadhyay2022_iontrap} proposed a heuristic mapping strategy that outperforms the greedy mapping in \citet{murali2020architecting_iontrap} by attenuating the weights of gates that occur later in the program. Across 120 random circuits, their method achieves an average 9\% reduction in shuttling operations per program (with a maximum reduction of 21.3\%) and enhances program fidelity by up to $3.3\times$ (with an average improvement of $1.41\times$). \citet{ovide2024scaling} introduced the Spatio-Temporal Aware (STA) mapping algorithm for linear and ring topologies in QCCD architectures, building on prior work\cite{murali2020architecting_iontrap, saki2022muzzle}. The STA algorithm strategically positions qubits with stronger spatio-temporal correlations closer together, leveraging the temporal distribution of qubit interactions and their interaction ratios to optimize initial qubit placement. Compared to previous methods, the STA algorithm achieves significant improvements in execution efficiency, reducing total circuit execution time by up to 50\% across various benchmarks~\cite{ovide2024scaling}.

\citet{dai2024advanced} introduces an optimized shuttle strategy for parallel QCCD architectures, significantly reducing ion shuttle operations in trapped-ion quantum computing. The key contributions include a probabilistic formula for ion movement and novel methods for local layer generation and compression, which together enhance efficiency and fidelity while minimizing traffic blocks and trap capacity issues. Simulations demonstrate superior performance over existing strategies, highlighting the potential to accelerate quantum computation and improve scalability in NISQ devices. 

\citet{schmale2022backend_iontrap} propose a compiler designed to translate high-level quantum programming languages into TIASM, an assembly language for trapped-ion systems. Their approach leverages existing Python-based tools, such as Pytket~\cite{Sivarajah_2020}, for quantum gate conversion and gate-level circuit optimization. However, they further optimize circuit conversion for the TI machine, building on the framework from \citet{maslov2017basic}. Furthermore, they introduce heuristic-based strategies to optimize ion movement on the chip, including JunctionDistance, ComputeOrder, SpamStorage, and PartnerSorting. Under the optimal heuristic ion movement strategy, their approach reduces the number of movement operations by more than 50\% compared to randomly executed gates.

\citet{durandau2022automated_iontrap} introduce various techniques for initial qubit allocation and describes a method for fully automated shuttling. \textit{Order as is} (OAI) is the first initial ordering, in which qubit indexes are assigned from left to right according to the initial ion position in the segmented trap. The second strategy is called \textit{order inputs randomly} (OIR), where a uniform distribution is used for random assignment. The greedy approach of \textit{increase pairwise order} (IPO) is the last strategy. The heuristic used by the shuttle algorithm is the greedy strategy. The overall cost is mainly related to the time cost of the algorithm that calculates the shuttle sequence.

~\citet{kreppel2023quantum_iontrap} focus on the native gate set ($\mathcal{S}=\left\{\mathtt{R}\left(\theta,\phi\right),\mathtt{Rz}\left(\phi\right),\mathtt{ZZ}\left(\theta\right)\right\}$) and mainly optimizes the circuit transformation part of the compiler to achieve efficient translation of quantum circuits based on shuttle trapped-ions. For qubit mapping, they employed two Pytket techniques: GraphPlacement and LinePlacement. These mapping methods are based on heuristic methods. They then use Pytket's RoutingPass to insert exchange gates to limit the execution of two-qubit gates to ions with adjacent vertices. Regarding the shutting method, the method proposed by~\cite{durandau2022automated_iontrap} is used. Through the above process, the gate count can be reduced by up to $5.1\times$ and $2.2\times$ compared to traditional Pytket and Qiskit compilation, respectively.

\citet{zhu2025s} propose S-SYNC to improve the compilation performance of QCCD. Their method models the entire QCCD architecture as a graph, providing a unified and structured representation of the system. Within this framework, shuttle operations are abstracted as a specialized form of SWAP gate, thereby reformulating the circuit compilation task as a generic SWAP gate insertion problem. By considering both shuttle and SWAP operations, the approach facilitates the development of a scalable algorithm capable of efficiently co-optimizing ion movement and SWAP gate insertion.

\citet{bach2025efficient} formulate the junction part as a position graph, which facilitates the description of various details within the shuttle. Based on this setting, they proposed SHAPER and SHAW to demonstrate the usefulness of the position graph by transferring a state-of-the-art mapping algorithm from superconducting systems. They showed that these methods can resolve congestion and deadlocks and fully utilize two-dimensional QCCD-based architectures.

\citet{ruan2025trapsimdsimdawarecompileroptimization} propose a novel Single Instruction Multiple Data (SIMD) abstraction that captures the features of modular TI architectures and present FluxTrap, a SIMD-aware compiler framework that establishes a hardware–software co-design interface for TI systems.
FluxTrap incorporates two compilation passes: the SIMD aggregation pass, which merges scalar ion transport operations into wider SIMD instructions through a global search guided by a heuristic cost function, and the SIMD scheduling pass, which optimizes the selection and scheduling of these instructions under hardware constraints via hierarchical comparison and time-sliced synchronization.

In addition to the QCCD architecture discussed earlier, alternative QCCD designs feature fixed execution zones for gate operations. MQT introduces a configuration that represents a significant advancement over the traditional linear-tape structure~\cite{schoenberger2024using, schoenberger2024shuttling}. This new approach enables grid-like ion storage, offering greater flexibility compared to the linear-tape setup. However, unlike the standard QCCD architecture, this design mandates that all qubits be shuttled to a fixed execution zone for gate operations. To accommodate this setting, MQT provides two specialized compilers tailored for different QCCD configurations~\cite{schoenberger2024using, schoenberger2024shuttling}. The first compiler employs Boolean Satisfiability (SAT) to obtain exact solutions, optimizing the number of shuttles~\cite{schoenberger2024using}. The second compiler adopts a heuristic algorithm, leveraging a cycle-based approach that enhances scalability~\cite{schoenberger2024shuttling}.

\textbf{QEC compilers.} Compared to the optimization and implementation of quantum error correction codes in superconducting devices, research on trapped-ion systems remains relatively limited and less focused. \citet{leblond2023tiscc} develop TISCC, a software tool that generates circuits for a universal set of surface code patch operations in terms of a native trapped-ion gate set. TISCC is the first to translate the shuttle operation in QCCD into a lattice surgery operation that can be practically implemented, which enables the implementation of QEC based compilation in trapped-ion devices. \citet{yin2025flexion} then propose Flexion,  a comprehensive framework tailored to trapped-ion quantum computers that enables early fault tolerance by applying QEC selectively in space and time. Flexion employs bare qubits for single-qubit gates and QEC-encoded logical qubits for two-qubit gates. It develops a Runtime Encoding Protocol to facilitate conversion between these qubit types and integrates a Hybrid Instruction Set Architecture (ISA) and a dynamic encoding compiler to enhance the efficiency of quantum operations.

\subsection{Large Scale Distributed Device}
\label{subsec:large scale trapped-ion}
While the QCCD configuration allows for the design of large-scale quantum chips, it is limited by the capacity of individual traps. This limitation can lead to excessive shuttling in large-scale QCCDs, resulting in decreased fidelity. 
Consequently, distributed trapped-ion devices have been proposed by some scholars. TITAN \cite{Chu2024titan} is a recently proposed trapped-ion device that demonstrates significant potential for advancing the field.
This work clarifies the need for a distributed architecture as well as the difficulties in interacting with QCCDs through photonic switches and quantum material links. 

In addition, it explores the complexities of building distributed TI NISQ computers, highlighting the challenges and possible approaches to overcome existing technological limitations. It also includes a thorough examination of the conditions necessary to realize a scalable quantum computing architecture, emphasizing the critical function that distributed systems play in the development of quantum computing models. 
The key innovation presented in this paper is the use of photonic switches to enable interactions between ions that are located on different QCCD chips. The large-scale development of trapped-ion computers is made possible by this design paradigm, which marks a major advancement in the field of quantum computing. 
In addition, it presents an advanced partitioning and mapping algorithm, which minimizes inter-module communications and inter-QCCD communications within each TI module. Their results show that TITAN greatly enhances quantum application performance fidelity compared to existing systems.

\section{Neutral Atom Compiler}
\label{sec:neutral atom}
\subsection{Background}\label{sec:neutral_atom_background}

In neutral atom quantum computing, exciting atoms to high principal quantum number states (Rydberg states) is essential for realizing high-fidelity quantum gate operations~\cite{saffman2010quantum,henriet2020quantum}. Rydberg atoms exhibit dramatically enhanced interaction strengths compared to their ground-state counterparts due to the scaling of their dipole moments with the fourth power of the principal quantum number \( n \) and the radiative lifetime increasing as \( n^3 \)~\cite{wu2021concise}. 
This pronounced amplification enables the establishment of long-range, controllable dipole–dipole interactions between atoms, thereby providing the physical basis for two-qubit logic gates and entanglement generation. Notably, when one atom is excited to a Rydberg state, its strong interaction induces a significant energy shift in neighboring atoms, preventing their simultaneous excitation. This is a phenomenon known as the Rydberg blockade effect~\cite{jaksch2000fast, you2000quantum, urban2009observation}. The blockade mechanism ensures the single-excitation condition during quantum gate operations and is a critical technique for achieving deterministic entanglement and high-fidelity quantum gates.

Recent advances in optical tweezers technology have made it possible to precisely capture, manipulate, and rearrange individual neutral atoms~\cite{schlosser2001sub,kaufman2021quantum}. Using high numerical aperture lenses and holographic modulators, researchers can construct highly ordered atomic arrays in two or three dimensions~\cite{nogrette2014single, barredo2018synthetic}. 
Furthermore, programmable moving optical tweezers allow for real-time rearrangement of atoms, ensuring high filling factors and uniformity within quantum registers~\cite{endres2016atom, barredo2016atom,sheng2021efficient}. This technological breakthrough provides a strong foundation for the development of programmable Rydberg quantum computers~\cite{bluvstein2022quantum, bluvstein2024logical}, as it enables effective use of strong dipole interactions and blockade effects, which are realized by exciting atoms to Rydberg states within carefully engineered atomic arrays. Consequently, these advances have significantly propelled the implementation of quantum gate operations, entanglement generation, and large-scale quantum processing. Neutral atom quantum computers have demonstrated immense promise due to their exceptionally long coherence times, scalability, native multi-qubit gates, higher connectivity resulting from longer-range interactions, and the ability to produce identical and well-characterized qubits~\cite{morgado2021quantum_atom, henriet2020quantum_atom, evered2023high}. Up to 289 neutral atom qubits have been operated by~\citet{ebadi2022quantum}, and large system size increases are anticipated to continue~\cite{manetsch2024tweezerarray6100highly}.

Neutral Atom mainly has two models, one is from \citet{bluvstein2022quantum_atom} which a stationary, globally illuminating Rydberg laser is held fixed, while optical tweezer arrays dynamically rearrange individual atoms. By shortening the inter-atomic separation, atom pairs are brought into the Rydberg blockade regime, allowing the global laser to generate many pairwise entangled states in parallel. The other is from \cite{graham2022multi_atom} which the atoms remain fixed, and tightly focused, independently addressable beams are steered to selected sites. Local Rydberg excitation transiently enlarges the blockade radius of the targeted atoms, mediating controlled interactions only with their nearest neighbors and thereby realizing programmable, site-selective two-qubit entangling gates. 

\begin{table}[t]
\caption{A summary of Neutral Atom compilers in literature.}
\resizebox{\linewidth}{!}{
\renewcommand{\arraystretch}{1.5}
\begin{tabular}{c|c|c|c|c|c|c }
\toprule
% Long-distance interactions
\textbf{Ref.}& \textbf{ Year} & \textbf{Architecture Model} & \textbf{Problem} & \textbf{Algorithm} & \textbf{Atom Loss} &
\textbf{Movement} \\ \midrule
~\cite{baker2021exploiting_atom}& 2021 & 2D grid &   \textbf{M\&S} & Heuristic & \checkmark & \ding{55}  \\
 
~\cite{patel2022geyser_atom}& 2022 & 2D triangular grid & Circuit Optimization  & Heuristic &  \ding{55} &  \ding{55} \\
 
~\cite{tan2022qubit_atom}& 2022& 2D square grid RAA & \textbf{M\&S}  &  SMT &  \ding{55} & \checkmark \\
 
~\cite{li2023timing_atom}& 2023   & 2D grid & \textbf{M\&S} & Heuristic+ MCTS &  \ding{55} &  \ding{55}   \\
 
~\cite{tan2023compiling_atom}& 2023 & 2D square grid RAA & \textbf{M\&S}  & SMT + Heuristic &  \ding{55} & \checkmark  \\
 
\cite{nottingham2023decomposing_atom}& 2023&  2D square grid RAA& Circuit Decomposition + \textbf{M\&S} & Heuristic&  \ding{55} & \checkmark   \\
 
\cite{tan2023compiling_atom}& 2023& Zoned Architecture & \textbf{M\&S} & Heuristic &  \ding{55} & \checkmark  \\
 
\cite{wang2024atomique}& 2024& Zoned Architecture& \textbf{M\&S} & Heuristic &  \ding{55} & \checkmark   \\
 
\cite{lin2024reuse}& 2024& Zoned Architecture & \textbf{M\&S} & Heuristic &  \ding{55} & \checkmark   \\
 
\cite{ruan2024powermove} & 2024 & Zoned Architecture & \textbf{M\&S} & Heuristic &  \ding{55} & \checkmark  
\\
\midrule

% \bottomrule
\end{tabular}
}
\label{tab:atom-summarizations}
\end{table}

Neutral Atom model has several types of error. First, gate errors occur from imperfections in the laser or microwave pulses used to implement gates (e.g., variations in Rabi frequency) or from atomic decay processes due to the finite lifetime of the Rydberg state or off-resonant scattering from the intermediate state. The probabilities of drawing an error on a single-qubit and two-qubit gate are given, respectively, by $p_{1k}$ and $p_{2k}$ for $k \in \{X,Y,Z\}$, and probabilities of drawing no error are $1-\sum p_{1k}$ and $1-\sum p_{2k}$, where $p_{1x} = p_{1y} = 0.1\%$, $p_{1z}=0.4\%$, $p_{2x}=p_{2y}=0.2\%$ and $p_{2z}=0.5\%$ \cite{bluvstein2022quantum_atom}. Also, Idle errors occur from unwanted interactions with the environment causing qubit states to decoherence. Additionally, owing to constraints in hardware, atom loss is present. Atom loss refers to the phenomenon in which neutral atoms may be lost during or between calculations when forming sparse grids. The reason for atom loss is that atoms in neutral atomic systems are limited by optical dipole traps (such as optical tweezers). Although this method provides considerable flexibility in array geometries and the potential for scaling to accommodate a large number of atoms, the trapping potential per atom is relatively weak compared to trapped-ion architectures.

Neutral atom devices offer a distinct native gate set compared to superconducting and trapped-ion platforms, including support for native multi-qubit gates such as the CCZ gate. The study by \citet{nottingham2023decomposing_atom} introduces a compilation framework for decomposing quantum gates into the native gate set of neutral atom systems. Furthermore, the work investigates the role of native multi-qubit gates in circuit compilation, analyzing trade-offs associated with their use, particularly in relation to three-qubit gate implementations. 

In the following, we provide a comprehensive overview of state-of-the-art qubit mapping and routing compilers specifically designed for neutral atom architectures. The essential findings and a concise summary of the relevant information are provided in Table \ref{tab:atom-summarizations}. An in-depth analysis of these compilers will be elaborated upon in the following sections.

\subsection{Fixed Neutral Atom Arrays}
\label{subsec: early na}
We first discuss the early compilation work on neutral atom arrays. The term ``early'' refers to two aspects: firstly, the early development of neutral atom devices, which predominantly focused on fixed atom arrays (FAA) that lack the ability to move atoms. Secondly, it pertains to the nascent stage of compiler development for these systems. During this period, studies often assumed the use of individually addressable Rydberg entangling gates, leading to the adoption of design principles similar to those employed in superconducting quantum devices for compiler development.

Although the current technological implementations of individually addressed Rydberg gates require a distinct optical setup, this setup has not yet achieved a fidelity competitive with that of global Rydberg lasers. Specifically, the fidelity of individually addressed Rydberg gates stands at 92.5\% as reported by \citet{graham2022multi_atom}, compared to the 99.5\% fidelity of global Rydberg lasers as noted by \citet{evered2023high}. Despite this disparity, the compilation methods employed in these implementations provide valuable insights into the intricacies of neural atom compiler design. This understanding is crucial for the development of more effective compilers. Moreover, future technological advancements may enhance the viability and potential advantages of individually addressable Rydberg entangling gates, suggesting that these compilers may become increasingly significant in the field.

\textbf{Constraints. } In fixed atom arrays, qubit connectivity is constrained by the Rydberg blockade radius, which requires that the distance between any two qubits involved in a two-qubit gate be less than this radius. Fixed atom arrays also support the parallel execution of gates provided that the gates operate on the separate groups of qubits and the distances between qubits involved in different gates exceed the Rydberg blockade radius.

\textbf{FAA Compiler. }
\citet{baker2021exploiting_atom} propose the first compiler to account for hardware-level characteristics specific to neutral atom devices, including long-distance interactions and atom loss and taking in account several unique properties on neutral atomic devices, such as interaction distance, induced restriction zones, and neutral atom device tailored multi-qubit gates. 
They adapt an existing look-ahead heuristic~\cite{wille2016look} and incorporated Qiskit's optimization techniques to account for the specific constraints present in neutral atom devices. Furthermore, \citet{baker2021exploiting_atom} address the impact of atom loss, which poses a critical challenge for program execution. If an atom involved in the computation is lost, the result becomes incomplete, and it is often impossible to determine whether the loss occurred during execution. In such cases, the faulty run can be discarded and another interaction with the device initiated. To handle atom loss, one may recompile the program for the resulting sparser qubit layout, reload the atom array, or adapt the compiled circuit to account for the missing atom. While the first two approaches incur significant time overhead, the third offers a faster alternative that enables more executions within the same time frame while preserving program validity. 
\citet{baker2021exploiting_atom} support this observation by modeling atom loss and demonstrating that it can be detected through fluorescence imaging. They further explore several software and hardware remedies, each with varying overhead costs, to mitigate the impact of such loss.

GEYSER \cite{patel2022geyser_atom} is a compilation framework for Quantum Computing with neutral atoms that aims to optimize gate usage by combining gates applied to the same qubit block. It leverages the native multi-qubit gates of neutral atoms, such as the CCZ gate, which requires significantly fewer pulses compared to decomposing it into CZ gates. The framework focuses on 3-qubit blocks, as extending to higher-qubit gates (e.g., CCCZ) is more challenging and results in larger restriction zones. GEYSER consists of three steps: (1) Circuit mapping to a triangular topology using Qiskit \cite{aleksandrowicz2019qiskit}, (2) Circuit blocking to generate triangular blocks through a heuristic function, despite the large search space, and (3) Block composition using parameterized quantum circuits (PQCs) to approximate the blocks, although a general decomposition for 3-qubit blocks remains elusive. Overall, GEYSER demonstrates potential for reducing gate counts in quantum circuits.

TETRIS \cite{li2023timing_atom} is an abstract model of the \textbf{M\&S} problem that considers the duration differences of gates and other NA characteristics. For physical qubits arranged in an $n$-dimensional space, the model represents the space as $n$-dimensional. The height of the constructed blocks corresponds to the execution time of the given circuit. Thus, the objective is to minimize this height by carefully scheduling the gates.
The paper proposes two algorithms for solving the TETRIS problem. The first is the Heuristic Greedy Algorithm (HGA), which quickly schedules gates based on heuristic observations. The second is an improved Monte Carlo Tree Search (MCTS) algorithm. Compared to LAC \cite{baker2021exploiting_atom}, HGA demonstrates faster compilation times, while MCTS achieves even better performance, especially for large circuits. However, MCTS is slower than the other two algorithms due to its extensive exploration of potential states.

\subsection{Reconfigurable Neutral Atom Arrays}
\label{subsec: adavanced na}
\begin{figure}[h]
    \centering
    \includegraphics[width=0.6\linewidth]{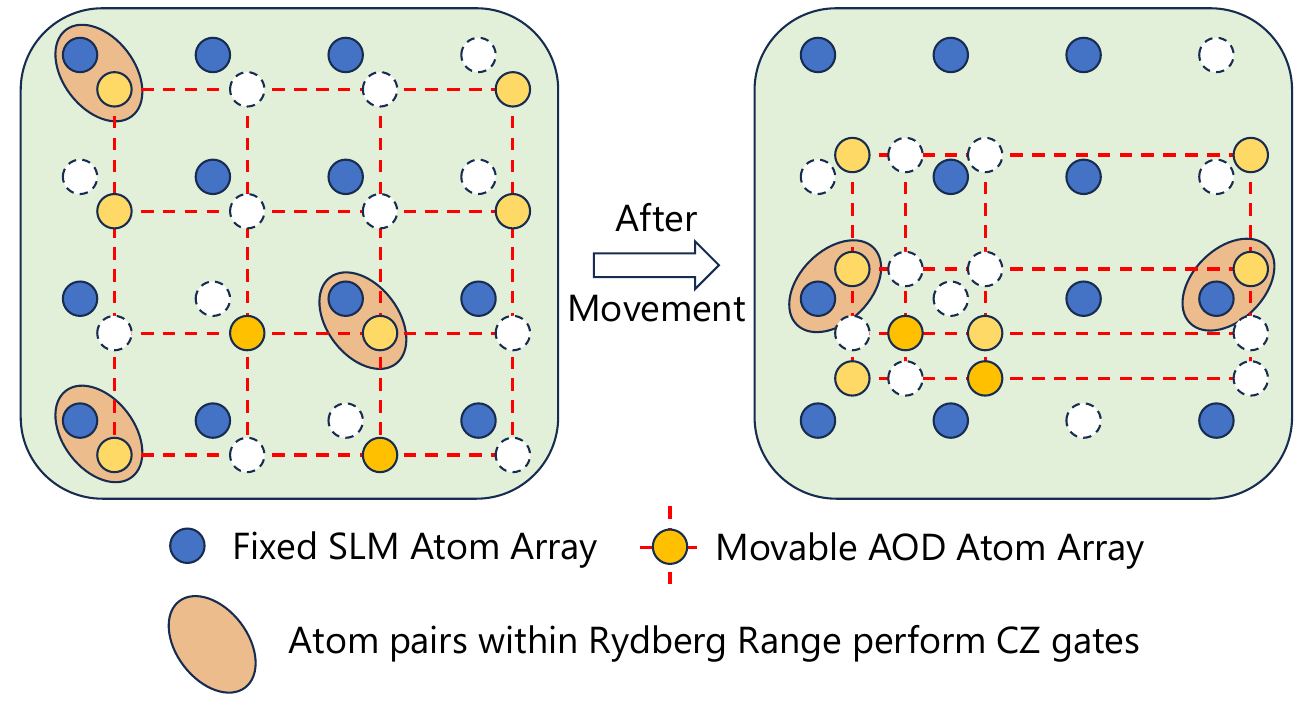}
    \caption{ RAA with fixed and movable atoms. It is important to note that while the AOD atom array is movable, the movement is collective in nature; the relative positions of the AOD atoms within the array remain unchanged. Furthermore, due to the current neutral atom devices' limited individual addressing capabilities, all pairs within the Rydberg range are required to perform compulsory two-qubit gates simultaneously.}
    \label{fig:RAA example}
\end{figure}

One advantage of neutral atoms is the ability to physically move qubits from one location to another during execution of the circuit. Devices that enable such qubit movement are commonly known as field programmable qubit array (FPQA) or Reconfigurable Atom Arrays (RAA) \cite{bluvstein2022quantum_atom,bluvstein2024logical}, which is shown in Fig.~\ref{fig:RAA example}.

\textbf{Constraints. }
Although RAA devices share certain limitations with FAA devices, they also face additional constraints. In RAA systems, each atom in the array serves as a qubit and is held in an optical trap. Some of these traps, generated by a spatial light modulator (SLM), remain stationary and function like quantum registers in a fixed architecture. In contrast, atoms held in traps created by a two-dimensional acousto-optic deflector (AOD) are movable. 

For routing with atom movements, there are four main constraints. 
(1) When moving from one location to another, an atom cannot get within some threshold distance $d_{thr}$ from any other (stationary or moving) qubits. 
(2) If one label atoms’ starting positions within the 2D array with coordinates $(x, y)$ and ending coordinates $(x' , y' )$, simultaneous movement operations can occur on atoms with the same x value only if it will also result in the same $x'$ value. 
(3) Similarly, one can simultaneously move atoms with the same y value only if they end at the same $y'$ value. 
(4) AOD row/col can not cross. 

Despite these limitations, qubit movement can outperform SWAP gates in certain scenarios. 
For instance, implementing a SWAP gate using native operations typically requires 3 CZ gates, 8 GR gates, and 10 $R_z$ gates, making it highly resource-intensive. Several results have shown that atom movement offers better performance in terms of both fidelity and circuit duration compared to SWAP-based approaches~\cite{tan2022qubit_atom, nottingham2023decomposing_atom}.

\textbf{RAA Compiler.}
\citet{bluvstein2022quantum_atom} show experimentally that entangled qubits can be coherently transported across a two-dimensional atom array with no negative effect on qubit entanglement and fidelity when qubits are moved at a speed of 0.55 \textmu  m/\textmu  s or slower. 
The advanced neural atom compiler incorporates a broader spectrum of characteristics from neutral atomic devices into its considerations. Using these intrinsic properties, they achieve superior compilation results.

OLSQ \cite{tan2022qubit_atom} is the first to take atom movement into account. Compare with the traditional \textbf{M\&S} solution, they use atom movement to meet the coupling instead of inserting swap gates. They formulate the \textbf{M\&S} problem in neutral atom device to a satisfiability modulo theories (SMT) model. The total number of constraints is $O(G^2 +GT + N^2T)$, where $T$ is the number of stages (Stages are segmented by movement), $N$ is the number of qubits and $G$ is the number of gates. The worst-case runtime of SMT solving is exponential, i.e., $O((N_{SLM}N_{AOD})^{NT}\cdot T^G)$ where $N_{SLM}= XY$ is the total number of SLM traps, and $N_{AOD}$ is the total number of AOD traps. Therefore, such method only work well on shallow and simple circuit.

\citet{nottingham2023decomposing_atom} propose a routing algorithm that assumes that AOD rows and columns can cross–specifically when two atoms are simultaneously moving in opposite directions without breaking these constraints. Their experiment result shows, atom movement works better both on fidelity and circuit duration than swap-based method.

Q-Pilot \cite{wang2023q} introduces a compilation strategy inspired by FPGA placement and routing techniques. All data qubits are mapped to fixed atoms, while movable atoms serve as ancilla qubits to facilitate routing between data qubits. These movable ancilla qubits, termed ``flying ancillas'', are dynamically generated and recycled during execution to perform two-qubit gates. To enhance parallelism and reduce circuit depth, a high-parallelism generic router is implemented. This router dynamically arranges movable AOD atoms and schedules two-qubit gates. It employs a heuristic-based approach to select as many parallelizable gates as possible within the constraints of the FPQA architecture.

Atomique \cite{wang2024atomique} develops compilation framework that addresses the unique challenges of mapping qubits, moving atoms, and scheduling gates in RAAs. For inter-array entangling gates, where one qubit is in the SLM array and the other is in the AOD array, the entire AOD array is moved to bring the qubit pairs into proximity for interaction.
The framework includes a qubit-array mapper that uses a MAX $k$-Cut algorithm on a constructed gate frequency graph to determine the coarse-grained mapping of qubits to arrays, aiming to minimize SWAP operations by maximizing inter-array two-qubit gates. Additionally, a qubit-atom mapper focuses on fine-grained mapping of qubits to specific atoms within the arrays, considering load balancing to prevent hardware constraint violations and ensure efficient parallel gate execution. Finally, a high-parallelism router schedules the movements of AOD rows and columns and the execution of gates, identifying parallel gates that can be executed simultaneously while adhering to hardware constraints such as the Rydberg blockade mechanism and the preservation of row/column order.

Enola \cite{tan2024compilation} introduces a novel approach to quantum circuit compilation by leveraging graph theory. The process begins with the scheduling of entangling gates using an edge-coloring algorithm, which effectively minimizes the number of layers required for the quantum circuit. This step ensures that the gates are arranged in a near-optimal manner. 
They prove that for a group of commutable two-qubit gates on  $n$ qubits and it has $O(n^2)$ gates, there exists an algorithm with time complexity $O(n^3)$ that assigns these gates to at most $S_{opt}$ + 1 Rydberg stages, where $S_{opt}$ is the optimal number of stages.
Following the scheduling phase, Enola employs a maximal independent set algorithm to derive rounds of parallel qubit movements between the layers. This approach allows for efficient reconfiguration of qubits, ensuring that the movement of qubits between different layers is optimized for parallelism and reduced execution time. 

\subsection{Zoned Architecture Neural Atom Devices}
\label{subsec: zoned na}
\begin{figure}[h]
    \centering
    \includegraphics[width=0.4\linewidth]{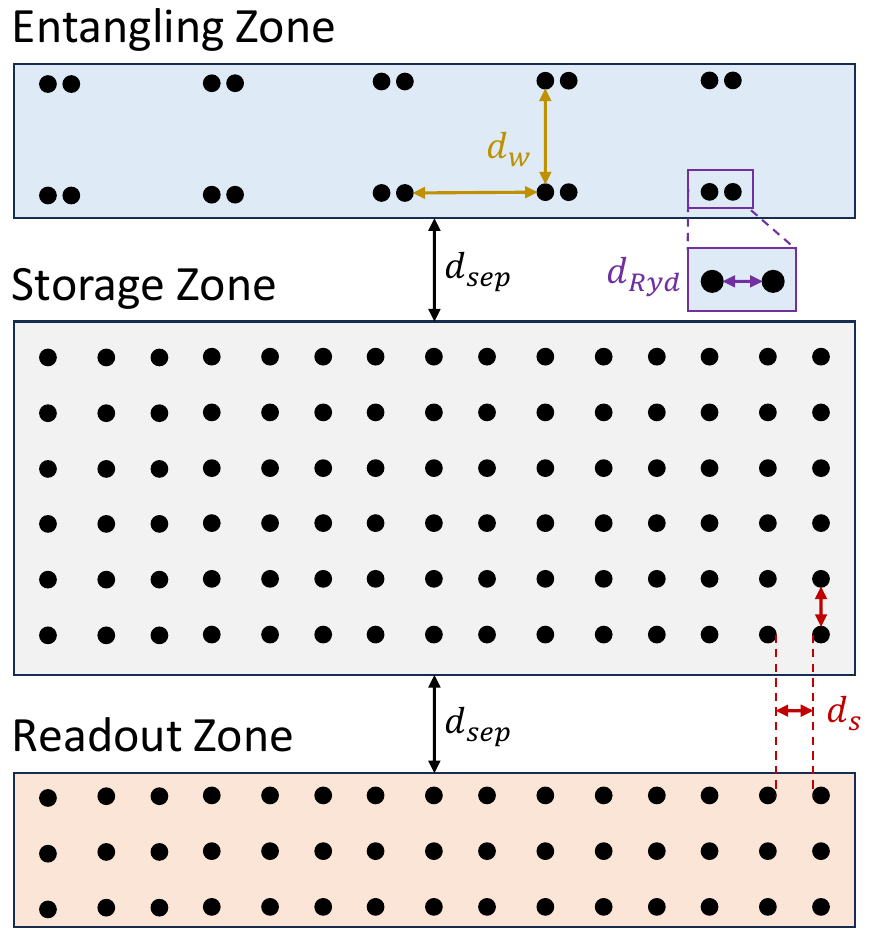}
    \caption{The typical zoned architecture, proposed by \citet{decker2024arctic}, is composed of three distinct regions: a readout zone (colored orange) designated for qubit measurement, an entanglement zone (colored blue) where Rydberg entangling gates are executed, and a storage zone (colored grey) where single-qubit gates are executed.
    The number of rows and columns in each zone is variable, allowing for flexibility in the architecture's configuration. The distance between zones, denoted as \(d_{sep}\), ranges from 10 \cite{lin2024reuse} to 20 \textmu m. The spacing between qubits when they are in the storage zone or readout zone, represented as \(d_{s}\), varies between 2 to 6 \textmu m. The distance at which two qubits can become entangled, \(d_{Ryd}\), is set at 2 \textmu m. Lastly, the separation between qubit pairs within the entangling zone, \(d_{w}\), is specified as 10 \textmu m.
    }
    \label{fig:zoned architecture NA device}
\end{figure}

Recent advancements in quantum computing have led to the emergence of zoned architectures, which offer promising enhancements in algorithmic performance and fidelity. The structure of such devices is shown as Fig.~\ref{fig:zoned architecture NA device}. Within this framework, the movement of atoms within the zone architecture is also facilitated by the implementation of AOD.
While it has demonstrated improvements in fidelity compared to monolithic architectures, its applicability is somewhat limited due to the constraints imposed by this particular configuration.

\textbf{Constraints.} 
The overall constraints of zoned architecture NA devices are consistent with those of RAA devices. However, they can only apply Rydberg entangling gates within the entanglement zone, while single-qubit gates can be applied within the storage zone.

\textbf{Zoned Architecture NA Compiler.}
NALAC~\cite{stade2024abstract} is designed to compile quantum circuits for zoned architectures of various sizes. It achieves qubit reuse by moving two rows of qubits from the storage zone to the entangling zone and employing a ``sliding'' mechanism to facilitate interactions between qubits. However, the tool's greedy placement strategy, which prioritizes minimizing movement within a single stage, restricts gate placement to a single row in the entangling zone. Furthermore, the current limitations of NALAC, particularly its lack of support for architectures featuring multiple AODs and a variety of zone configurations, impede its capacity to fully explore and adapt to more sophisticated and advanced system designs.

ZAC~\cite{lin2024reuse} is a novel compiler for zoned quantum architectures, featuring a reuse-aware placement strategy and a load-balancing scheduling mechanism. It is specifically designed for architectures equipped with multiple AODs and diverse zone configurations. By optimizing qubit reuse and efficiently distributing rearrangement jobs across AODs, ZAC minimizes data movement overhead and maximizes hardware utilization, thereby enhancing the performance of quantum circuit execution. Additionally, ZAC introduces ZAIR, a simplified intermediate representation for zoned architectures. ZAIR streamlines the compilation process by abstracting qubit movements into rearrangement jobs, making it easier to generate machine-level instructions and leverage multiple AODs for parallel qubit manipulation.

ZAP~\cite{huang2024zap} is a compiler shares some similarities with ZAC but without an iterative qubit mapping strategy.  ZAP employs two different ASAP (As Soon As Possible) scheduling strategies combined with a non-iterative placement scheme, significantly improve circuit fidelity and scalability, and largely reducing the compile runtime with two orders magnitude improvement. It also minimizes atom movements, and thereby reducing the crosstalk errors by reusing atoms in the entanglement zone.

PowerMove~\cite{ruan2024powermove} enhances the qubit movement framework while fully integrating the advantages of zoned architecture. 
Their solution consists of three key components:
The ``Stage Scheduler'' optimizes stage execution order to minimize qubit interchange between computation and storage zones during layout transitions, reducing inter-zone movement overhead.
The ``Continuous Router'' integrates qubit allocation and movement for seamless transitions between desired layouts without fixed intermediates.
The ``Coll-Move Scheduler'' optimizes collective qubit movements to maximize storage zone dwell time, minimizing decoherence, and enhances movement parallelism by scheduling multiple AOD arrays.

\section{Future Direction and Conclusion}\label{sec:conclusion}

This study offers a comprehensive analysis of the latest advancements in compiler design, focusing on three distinct quantum computing platforms: superconducting devices, trapped-ion devices, and neutral atom devices. We begin by providing an overview of the essential background knowledge, including an introduction to quantum compiler design and relevant fundamental concepts. Our analysis indicates that compiler research for superconducting devices is relatively well-developed, while studies targeting trapped-ion and neutral atom platforms remain comparatively limited. Given the current landscape, there is significant opportunity for further exploration and advancement in the design of quantum compilers, particularly those tailored to the unique constraints of emerging hardware architectures. In the following, we outline several promising directions for further investigation.

\textbf{AI for Compiler Design.} Artificial intelligence techniques have become indispensable across science and engineering, with deep learning and reinforcement learning now matching or surpassing human-designed heuristics and achieving state-of-the-art accuracy in tasks such as protein folding~\cite{jumper2021highly} and decoding noisy quantum syndromes~\cite{bausch2024learning}. Building on these advances, future quantum compiler research can move beyond handcrafted cost functions by enabling learning agents to infer optimal qubit mappings and routing strategies directly. A data-driven compiler could autonomously uncover subtle spatio-temporal patterns, such as correlations between connectivity and noise across different hardware that are often overlooked by traditional heuristics. Nevertheless, applying AI to compiler design presents several open challenges. First, learned policies must be provably correct, ensuring that the generated mappings and schedules consistently adhere to hardware connectivity constraints. Achieving such correctness will likely require hybrid approaches that incorporate formal verification or rule-based safeguards~\cite{ren2024leveraging} within the inference process. Additional challenges include generating representative training data for emerging hardware platforms, enabling sample-efficient exploration of the vast configuration space, and producing interpretable outputs that offer meaningful insights into the compiler's decision-making process.

\textbf{Quantum Application-Specific Compilers.} Application-specific quantum compilers are becoming increasingly important as the field moves from general-purpose demonstrations toward targeted use cases such as Hamiltonian simulation. By tailoring qubit mappings and routing strategies to the structural regularities of a given algorithm, these compilers can substantially reduce circuit depth and show improvements that are difficult to achieve with generic compilers. Moreover, domain-aware optimizations often expose hardware-friendly patterns that can be exploited to boost fidelity on current noisy devices or to minimize overhead. Developing a broader suite of application-specific compilers represents a compelling direction for future research. Currently, most algorithm-specific compiler efforts focus primarily on superconducting devices, limiting their applicability across diverse quantum platforms. As quantum hardware technologies continue to evolve, with emerging functionalities such as mid-circuit measurement now supported~\cite{corcoles2021exploiting, pino2021demonstration, paler2016wire, hua2023caqr, brandhofer2023optimal, fang2023dynamic, uchehara2024graph}, it becomes increasingly important to incorporate these hardware specific constraints into the design of quantum compilers. Addressing these complexities requires an integrated approach to algorithm-software-hardware co-design that can effectively leverage the unique capabilities of each platform and enable more efficient, high fidelity quantum computation.

\textbf{Quantum Error Correction Compilers.} As the field progresses toward fully fault-tolerant quantum computation, it is entering an intermediate stage that bridges today’s NISQ processors and future large-scale error-corrected systems, referred to as FASQ (Fault-Tolerant Application-Scale Quantum) computers~\cite{preskill2025beyond}.
With devices now offering larger qubit counts and higher fidelities, partial quantum error correction~\cite{koukoulekidis2023frameworkPEC, bultrini2023battle} is becoming practical, prompting a reassessment of the computational tasks these intermediate systems can undertake. Consequently, compiler design must evolve from heuristic NISQ mapping to FASQ-oriented pipelines capable of scheduling tens of thousands of QEC-protected gates while controlling space–time overhead. Yet most current compilers remain focused on NISQ architectures and thus neglect the challenges of deploying QEC codes on contemporary hardware. Moreover, the rapid diversification of quantum platforms—superconducting circuits, trapped-ions, neutral atom arrays and others—demands QEC-aware compilation frameworks that explicitly account for platform-specific constraints such as native gate sets, connectivity graphs, transport costs, and decoder latency. Most existing compiler frameworks focus on implementing QEC on superconducting platforms~\cite{wu2022synthesis, yin2025qecc, yin2024surf, fang2024caliscalpel} and therefore cannot be directly applied to architectures with different physical constraints. Developing truly hardware-aware, fault-tolerant compilers that account for each platform’s native gate set, connectivity, and noise profile will not only maximize the performance of near-term FASQ processors but also provide essential guidance for selecting and adapting QEC codes to emerging quantum technologies.

\textbf{Developing Standardized Benchmarks.} Benchmark suites serve as standardized sets of tests designed to evaluate the performance of various software systems or computing platforms. Historically, as digital computing advanced, LINPACK~\cite{dongarra1979linpack, dongarra2003linpack} was introduced to assess the capabilities of supercomputers and workstations. With the rise of machine learning, benchmark suites such as MLPerf~\cite{reddi2020mlperfinferencebenchmark, mattson2020mlperftrainingbenchmark, mattson2020mlperf} and GLUE~\cite{wang2019gluemultitaskbenchmarkanalysis} emerged to evaluate the performance of machine learning models and natural language understanding systems, respectively. In a similar way, quantum compiler design also necessitates standardized benchmarks~\cite{zhu2025quest}. Several benchmark collections~\cite{nation2025benchmarking} have been proposed to evaluate quantum circuit compilation performance, providing a foundation for consistent evaluation. However, current benchmark suites lack cross-platform comparability, as differences in hardware characteristics such as coherence time and connectivity make fair comparisons challenging. Developing standardized, platform-aware benchmarks is essential for enabling meaningful comparisons and guiding the selection of quantum algorithms for the most suitable hardware platforms.

%%
%% The acknowledgments section is defined using the "acks" environment
%% (and NOT an unnumbered section). This ensures the proper
%% identification of the section in the article metadata, and the
%% consistent spelling of the heading.
\begin{acks}
We would like to thank Zhixin Song, Keming He, Qiang Zheng, Zhiding Liang for insightful discussions and helpful comments.
This work was partially supported by the National Key R\&D Program of China (Grant No.~2024YFB4504004),  the National Natural Science Foundation of China (Grant No.~12447107), the Guangdong Natural Science Foundation (Grant No.~2025A1515012834), the Guangdong Provincial Quantum Science Strategic Initiative (Grant No.~GDZX2403008, GDZX2403001), the Guangdong Provincial Key Lab of Integrated Communication, Sensing and Computation for Ubiquitous Internet of Things (Grant No.~2023B1212010007), and the Education Bureau of Guangzhou Municipality.
\end{acks}

%%
%% The next two lines define the bibliography style to be used, and
%% the bibliography file.
\bibliographystyle{ACM-Reference-Format}
\bibliography{ijcai23}

\end{document}